\documentclass[aps,prd,showpacs,preprintnumbers,eqsecnum,amsmath,amssymb,showpacs,showkeys,nofootinbib]{revtex4}
\usepackage{graphicx}
\usepackage{dcolumn}
\usepackage{bm}
\newcommand{\matrixel}[3]{\left <\, #1 \left |\, #2\,\right |\, #3\,\right >}

\begin{document}
\title{Forward-backward and CP-violating asymmetries in rare {\boldmath $B_{d,s}\to (V,\gamma)\, \ell^+\ell^-$} decays}
\author{Irina Balakireva$^{a}$, Dmitri Melikhov$^{a,b,c}$, Nikolai Nikitin$^{a}$, and Danila Tlisov$^{a}$} 
\affiliation{
$^a$SINP, Moscow State University, 119991, Moscow, Russia\\
$^b$HEPHY, Austrian Academy of Sciences, Nikolsdorfergasse 18, A-1050, Vienna, Austria\\
$^c$Faculty of Physics, University of Vienna, Boltzmanngasse 5, A-1090 Vienna, Austria}
\date{\today}
\begin{abstract}
We study the forward-backward and the $CP$-violating asymmetries (both time-independent and time-dependent)   
in rare semileptonic $B_d\to\rho^0\ell^+\ell^-$, $B_s\to\phi\ell^+\ell^-$ 
and radiative leptonic $B_{d,s}\to \gamma \ell^+\ell^-$ decays and investigate the sensitivity of these asymmetries 
to the extensions of the Standard model. 
\end{abstract}
\pacs{13.20.He, 11.80.Cr, 11.30.Qc}   
\keywords{CP Violation, Rare Decays}
\maketitle
\section{Introduction}
Rare $B$-decays induced by flavor-changing neutral currents 
provide a valuable possibility of an indirect search of physics 
beyond the Standard Model (SM): such decays are forbidden at the tree level in the SM 
and occur through loops which contain contributions of potential new particles. These   
contributions modify the effective Hamiltonian describing rare $B$-decays 
and influence, e.g., the differential distributions and $CP$-violating effects. 

$CP$-violation in beauty sector has been measured for the first time 
at $B$-factories BaBar and Belle in two-hadron $B$-meson decays \cite{CP-in-B}. 
Other interesting reactions, where $CP$-violating effects may be studied, 
are rare semileptonic and radiative leptonic decays induced by $b \to (d, s)\,(\gamma,\, \ell^+ \ell^-)$
transitions (see, e.g., \cite{stone} and refs therein). 

In the SM one expects the branching ratios at the level $10^{-6}\div 10^{-7}$ for rare semileptonic $B$-decays 
\cite{mns} and $10^{-8}\div 10^{-9}$ for radiative leptonic $B$-decays \cite{mk,mnt}.
Presently, the largest available samples of $B$ mesons are accumulated by $B$-factories  
BaBar and Belle: in the recent publications these collaborations have announced 
384 and 657 million $B\bar B$-pairs, respectively \cite{A_FB-experiment}. 
Samples of $\bar BB$-mesons of this size allow one not only to measure integrated partial rates of rare semileptonic and 
(potentially) radiative leptonic decays, but also to study differential distributions 
in $B \to (K^*, K) \mu^+ \mu^-$ with several hundred events at hand. 
The shape of one of these differential distributions---the lepton forward-backward 
asymmetry $A_{FB}(s)$ ($\sqrt{s}$ being the dilepton invariant mass)---is known to be sensitive to the 
extentions of the SM, in particular to the sign of the Wilson coefficient $C_{7\gamma}$ in the effective Hamiltonian for 
$b\to (d,s)$ transitions. 
Recent results for $A_{FB}$ from Belle seem to indicate the sign of $C_{7\gamma}$ opposite to the SM prediction; 
the BaBar results based on lower statistics do not contradict to the results from Belle \cite{A_FB-experiment}. 
Unfortunately, the decisive conclusions cannot be made 
because of the limited statistics: the ``disagreement'' between the measuremt and the SM-prediction for the asymmetry is 
at the level of two standard deviations.  

Interesting information on the extentions of the SM may be obtained also from other reactions. 
In this paper we focus on the CP-violating effects in rare semileptonic and radiative leptonic $B$-decays. 
 
Obviously, an experimental study of $CP$-violating observables requires greater samples of beauty hadrons than 
those provided by the B-factories: since $CP$-violation effects in beauty sector are of order $10^{-3}$, 
one needs $B$-meson data samples exceeding those from $B$-factories by at least two orders of magnitude. 
One expects such data samples of beauty particles at the LHC: e.g., the detector LHCb is expected 
to register about $10^{12}$ $b\bar b$-pairs per year, 
approximately by four orders of magnitude more than a yearly yield of $b\bar b$-pairs at the $B$-factories \cite{LHCb}. 
A widely discussed Super-$B$ factory project, having a lower yield of $b\bar b$-pairs compared to the LHC, will 
provide much better signal-to-background ratio \cite{SuperB}. 
Thus, the new accelerators open exciting possibilities of an experimental study of $CP$-violating effects 
in rare $B$-decays with two leptons in the final state. 

The issue of $CP$-violation in rare $b$-decays has been already discussd in the literature, but mainly 
in connection with rare radiative decays: the structure of the helicity amplitudes --- the main theoretical 
tool for the analysis of $CP$-violating effects  --- is very simple in this case. Therefore various extentions of the SM 
may be analysed. A high sensitivity of the time-dependent $CP$-asymmetry 
$A_{CP}(\tau)$ in $B \to (K^*, \rho) \gamma$ to left-right symmetric models has been reported \cite{soni}. 

The time-independent $CP$-asymmetries in rare semileptonic decays were considered first for inclusive $B$-decays, 
i.e., for the process $b \to d \ell^+ \ell^-$ \cite{krugerCP}. 
An asymmetry of the order of a few percent in $b \to d \ell^+ \ell^-$ has been predicted; 
for the $b \to s \ell^+ \ell^-$-transitions the asymmetry is expected to be much smaller. 
Following \cite{krugerCP}, the time-independent $CP$-asymmetries in exclusive $B$-decays for several extentions of the 
SM have been analyzed \cite{CPextra,hiller}.

The time-dependent CP-asymmetries \cite{dunietz} were studied for the case of rare semileptonic 
$B_{s}\to \phi \ell^+ \ell^-$--decays in Ref.~\cite{hiller}. 

The present work provides the analysis of the asymmetries (forward-backward, time-independent and 
time-dependent $CP$-violating asymmetries) in rare semileptonic and radiative leptonic 
$B_{s,d}\to (V,\gamma)  \ell^+ \ell^-$-decays. 
We make use of the technique of helicity amplitudes developed in \cite{hagiwara,soni}. 
Flavour oscillations of the decaying neutral $B$-mesons are taken into account; this effect is shown to 
influence considerably the resulting CP-violating asymmetries. 

The paper is organized as follows: 
In Section \ref{Sect:2} we recall the structure of the effective Hamiltonian and the relevant amplitudes describing 
rare $B$-decays. Sections \ref{Sect:3} and \ref{Sect:4} provide the definitions of the asymmetries and their expressions 
in terms of the helicity amplitudes.  
Section \ref{Sect:5} contains the results of the numerical analysis. Conclusion summarizes our findings. 
Appendices contain all necessary technical details related to the helicity amplitudes for rare 
semileptonic (Appendix B) and rare radiative leptonic (Appendix C) decays. 

\section{Effective Hamiltonian and form factors for rare semileptonic and radiative leptonic $B$-decays
\label{Sect:2}}
Let us briefly recall the basic definitions necessary for the calculation of the amplitudes 
of $B$-decays of interest. 

\subsection{The effective Hamiltonian}
The effective Hamiltonian describing the $b\to s \ell^+\ell^-$ transition in the SM has the 
form \cite{wc}\footnote{We use the following conventions: 
$\gamma^5=i\gamma^0\gamma^1\gamma^2\gamma^3$, 
$\sigma_{\mu\nu}=\frac{i}{2}[\gamma_{\mu},\gamma_{\nu}]$, 
$\varepsilon^{0123}=-1$, $e=\sqrt{4\pi\alpha_{\rm em}} > 0$.} 
\begin{eqnarray}
\label{SMb2qll}
H_{\rm eff}^{\rm SM}
(b\to s\ell^+\ell^-) &=& \,\frac{G_{F}}{\sqrt2}\, \frac{\alpha_{em}}{2\pi}\,
V_{tb}V^*_{ts}\,
\left[\,-2\, {\frac{C_{7\gamma}(\mu)}{q^2}} \cdot 
\bar s\, i\sigma_{\alpha\beta}q^\beta\left\{m_b(1+\gamma^5)+ m_s(1-\gamma^5) \right\} b
\cdot 
\bar \ell\gamma^{\alpha}\ell\right. \nonumber \\
&&\left. +\, 
C_{9V}^{\rm eff\,(s)}(\mu,\, q^2)\cdot \bar s\gamma_{\alpha}\left (1-\gamma^5\right ) b \cdot \,\bar \ell\gamma^{\alpha}\ell \, 
+\,
C_{10A}(\mu)\cdot \bar s\gamma_{\alpha}\left (1-\gamma^5\right ) b \cdot \,{\bar \ell}\gamma^{\alpha}\gamma^5\ell \right], 
\end{eqnarray} 
where $m_b$ ($m_s$) is the $b$ ($s$)--quark mass, $V_{ij}$ are the elements of the CKM matrix,  
$\mu$ is the renormalization scale, and $q$ is the momentum of the $\ell^+\ell^-$ pair. 
The corresponding expression for the case of the $b\to d$ transition is self-evident.

The Wilson coefficient $C_{9V}^{\rm eff}(\mu,\, q^2)$ contains the contributions of the virtual $\bar uu$ and $\bar cc$ pairs, which 
involve the integration over short and long distances. The long-distance effects are described by the 
neutral vector-meson resonances $\rho$, $\omega$, and $J/\psi$, $\psi'$. We make use of the parameterization 
of $C_{9V}^{\rm eff}(\mu,\, q^2)$ from \cite{krugerCP} where the resonance contributions are modeled in a 
gauge-invariant way \cite{melikhovplb}.

The effective Hamiltonian for the $\bar b\to \bar s\ell^+\ell^-$ transition is obtained from (\ref{SMb2qll}) 
by interchanging $b$ and $s$ fields, i.e., by replacing $V_{qb}V^*_{qs}\to V^*_{qb}V_{qs}$, 
$\bar s \to \bar b$, $b \to s$, $m_b\leftrightarrow m_s$, $q \to -q$.\footnote{Let us notice that 
the effective Hamiltonian for the $\bar b\to \bar s$ transition is {\it not} related to that for the $b\to s$ 
transition by the hermitian conjugate: namely, 
$H_{\rm eff}(b\to s)
\simeq 
V^*_{qs}V_{qb}
T\{
j_\mu(q\to s)W^\mu \,
\bar q O q \,
j^\dagger_\nu(q\to b)W^{\dagger\nu}\}$ and 
$H_{\rm eff}(\bar b\to \bar s)
\simeq 
V_{qs}V^*_{qb}
T\{
j^{\dagger}_\mu(q\to s)W^{\dagger\mu } \,
\bar q O q  \,
j_\nu(q\to b)W^{\nu}\}$, where $(\bar q O q)^\dagger=\bar q O q$. 
The hermitian conjugate of $H_{\rm eff}(b\to s)$ does not give $H_{\rm eff}(\bar b\to \bar s)$, 
since the hermitian conjugate of the $T$-product gives the anti-$T$-product.}

\subsection{Form factors for weak $B\to V$ decays}
The form factors for rare semileptonic transitions 
$\bar B(p_1,\, M_1)\to \bar V(p_2,\, M_2,\,\varepsilon)$ are defined in the standard way (see e.g. \cite{BDandK}):
\begin{eqnarray}
\label{B2V-ff}
&&\langle\bar V (p_2,\, M_2,\,\varepsilon)  \left |
    \bar q\,\gamma_{\mu}\, b
\right|\bar B (p_1,\, M_1) \rangle\, =\, 
\frac{2\, V(q^2)}{M_1\, +\, M_2}\, \epsilon_{\mu \nu \alpha \beta}\, 
\varepsilon^{*\,\nu}\, p^{\alpha}_1\, p^{\beta}_2; \nonumber \\
&&\langle\bar V (p_2,\, M_2,\,\varepsilon)  \left |
    \bar q\,\gamma_{\mu}\gamma^5\, b
\right|\bar B (p_1,\, M_1) \rangle\, =\,
i\,\varepsilon_{\mu}^* (M_1\, +\, M_2)\, A_1(q^2)\, -\nonumber \\ 
&&\qquad -\, i\, (\varepsilon^*\, p_1)\, (p_1\, +\, p_2)_{\mu}\, 
\frac{A_2(q^2)}{M_1\, +\, M_2}\, -\, 
i\, (\varepsilon^*\, p_1)\, q_{\mu}\,\frac{2\, M_2}{q^2}\, 
\left (A_3(q^2)\, -\, A_0(q^2) \right ); \nonumber \\
&&\langle\bar V (p_2,\, M_2,\,\varepsilon)  \left |
    \bar q\,\sigma_{\mu\nu}\, q^{\nu}\, b
\right|\bar B (p_1,\, M_1) \rangle\, =\, 
2\, i\, T_1(q^2)\, \epsilon_{\mu \nu \alpha \beta}\, 
\varepsilon^{*\,\nu}\, p^{\alpha}_1\, p^{\beta}_2; \\
&&\langle\bar V (p_2,\, M_2,\,\varepsilon)  \left |
    \bar q\,\sigma_{\mu\nu}\gamma^5\, q^{\nu}\, b
\right|\bar B (p_1,\, M_1) \rangle\, =  T_2(q^2)\, \left (
\varepsilon_{\mu}^*\, Pq\, -\, 
(\varepsilon^*\, q)\, P_{\mu}\right )\,+\, T_3(q^2)\, (\varepsilon^*\, q)\,\left (
q_{\mu}\, -\,\frac{q^2}{Pq}\,P_{\mu} 
\right ),  \nonumber
\end{eqnarray}
where $q = p_1 - p_2$, $P = p_1 + p_2$, $p_1^2=M_1^2$, $p_2^2=M_2^2$, $\varepsilon^*$ is the 
polarization vector of the final vector meson, 
$\varepsilon^*p_2=0$. The form factors satisfy the following conditions
\begin{eqnarray}
A_3(q^2)=\frac{M_1+M_2}{2M_2}A_1(q^2)-\frac{M_1-M_2}{2M_2}A_2(q^2),\qquad 
A_0(0)=A_3(0), \qquad T_1(0)=T_2(0). 
\end{eqnarray} 

\subsection{Form factors for weak $B\to \gamma$ decays}
The $B\to\gamma$ amplitudes are parametrized as follows \cite{mk}:
\begin{eqnarray}
\label{barff}
\left <
  \gamma (k,\,\epsilon)|\bar q\gamma_\mu\gamma_5 b|\bar B^0_q(p, M_1)
\right >
&=& i\, e\,\epsilon^*_{\alpha}\,
\left ( g_{\mu\alpha} (pk)-p_\alpha k_\mu \right )\,\frac{F_A(q^2)}{M_1},
\nonumber
\\
\left <
  \gamma(k,\,\epsilon)|\bar q\gamma_\mu b|\bar B^0_q(p, M_1)
\right >
&=&
e\,\epsilon^*_{\alpha}\,\epsilon_{\mu\alpha\xi\eta} p_{\xi}k_{\eta}\,
\frac{F_V(q^2)}{M_1},
\\
\left <
  \gamma(k,\,\epsilon)|\bar q\sigma_{\mu\nu}\gamma_5 b|\bar B^0_q(p, M_1)
\right >\, (p-k)_{\nu}
&=&
e\,\epsilon^*_{\alpha}\,\left[ g_{\mu\alpha}(pk)- p_{\alpha}k_{\mu}\right ]\,
F_{TA}(q^2),
\nonumber
\\
\left <
\gamma(k,\,\epsilon)|\bar q \sigma_{\mu\nu} b|\bar B^0_q(p, M_1)
\right >\, (p-k)_{\nu}
&=&
i\, e\,\epsilon^*_{\alpha}\epsilon_{\mu\alpha\xi\eta}p_{\xi}k_{\eta}\,
F_{TV}(q^2).
\nonumber
\end{eqnarray}
Here $q=p-k$, $k^2=0$, $p^2=M_1^2$.  

\section{Forward-backward asymmetry}
\label{Sect:3}
The lepton forward-backward asymmetry $A_{FB}(\hat s)$ is one of the differential distributions 
relatively stable with respect to QCD uncertainties and sensitive to the new physics 
effects \cite{mns2}. Therefore it has been extensively studied both theoretically and experimentally 
\cite{BDandK}. We make use of the following definition of $A_{FB}(\hat s)$ for $\bar B\to \bar f$ decays 
\begin{eqnarray}
\label{Afb}
A_{FB}(\hat s)\, =\,
\frac{\displaystyle\int_0^1 d\cos\theta\,\frac{d^2\Gamma (\bar B \to \bar f)}
               {d \hat s\, d\cos\theta}\, -\,
      \int_{-1}^0 d\cos\theta\,\frac{d^2\Gamma (\bar B \to \bar f)}
               {d \hat s\, d\cos\theta}
     }
     {\displaystyle\frac{d\Gamma (\bar B \to \bar f)}{d\, \hat s}
     },
\end{eqnarray}
where $f=V l^+ l^-$ for rare semileptonic decay and $f=\gamma l^+ l^-$ for rare radiative decay;  
$\hat s\equiv s/M_B^2$, $\sqrt{s}$ being the dilepton invariant mass. 
The asymmetry is calculated in the rest frame of the lepton pair, and the angle
$\theta$ is defined as the angle between the $\bar B$-meson 3-momentum and the 3-momentum of the outgoing 
negative-charged lepton, $l^-$.

Equivalently, for $B\to f$ one defines 
\begin{eqnarray}
\label{Afb1}
A_{FB}(\hat s)\, =\,
\frac{\displaystyle\int_0^1 d\cos\theta_+\,\frac{d^2\Gamma (B \to f)}
               {d \hat s\, d\cos\theta_+}\, -\,
      \int_{-1}^0 d\cos\theta_+\,\frac{d^2\Gamma (B \to f)}
               {d \hat s\, d\cos\theta_+}
     }
     {\displaystyle\frac{d\Gamma (B \to f)}{d\, \hat s}
     }
\end{eqnarray}
where $\theta_+$ is the angle between the $B$-meson 3-momentum and the 3-momentum of the outgoing 
positive-charged lepton, $l^+$. If CP-violating effects are neglected, both asymmetries (\ref{Afb}) 
and (\ref{Afb1}) are equal to each other.  

The decay rate may be expressed in a simple way via the helicity amplitudes describing 
the $B\to (V,\gamma) \ell^+\ell^-$ and $\bar B\to (\bar V,\gamma) \ell^+\ell^-$ decay:
\begin{eqnarray}
\label{dgamma}
{\Gamma}=\frac{1}{2}
\int\,\frac{d\Phi_3}{2\, M_B}
\sum_{\lambda_i\,\lambda_1\,\lambda_2}\,
\left (
\left |
 A^{(q)}_{\lambda_i\,\lambda_1\,\lambda_2}(\hat s,\,\cos\theta)
\right |^2 + 
\left | \bar A^{(q)}_{\lambda_i\,\lambda_1\,\lambda_2}(\hat s,\,\cos\theta)
\right |^2
\right)
\end{eqnarray}
Explicit expressions for the helicity amplitudes and the phase-space volume are 
given in Appendix B for $\bar B(B)\to \bar V(V) \ell^+\ell^-$ decay and in Appendix C for $\bar B(B)\to \gamma \ell^+\ell^-$. 
Summation in Eq.~(\ref{dgamma}) runs over all possible values of the helicity indices 
(i.e. $\lambda_{1,2}=L,R$ for fermions, $\lambda_{i=V}=\pm 1,0$ for vector mesons, 
and $\lambda_{i=\gamma}=\pm 1$ for the photon). 
Making use of these expressions it is straightforward to calculate, e.g., $d^2\Gamma/d\hat s \,d\cos\theta$. 

Our present calculation of the forward-backward asymmetry in the  
$B\to\gamma \ell^+\ell^-$ decays takes into account Bremsstrahlung effects 
(the expression for the corresponding amplitude can be found in \cite{mnt}). 
We checked that the previous results for $A_{FB}$ from \cite{ds,mk},
where Bremsstrahlung was not considered, is reproduced if the Bremsstrahlung contribution in our 
formulas is omitted. 

\section{CP-violating asymmetries\label{Sect:4}}
Time-dependent CP-violating asymmetry is defined in the $B$-meson rest frame as follows \cite{waldi}
\begin{equation}
\label{ACPtau-def}
A_{CP}^{B_q\to f}(\tau)\, =\,
\frac{\displaystyle\frac{d\Gamma (\bar B^0_q\to f)}{d\tau} -\displaystyle\frac{d\Gamma (B^0_q\to f)}{d\tau}}
     {\displaystyle\frac{d\Gamma (\bar B^0_q\to f)}{d\tau}+\displaystyle\frac{d\Gamma (B^0_q\to f)}{d\tau}}, 
\end{equation}
where $f$ is the common final state for $B^0$ and $\bar B^0$ decays. In this case a pronounced CP violation is expected in 
interference between the oscillation and decay amplitudes \cite{sanda}. 
For instance, for leptonic radiative decays $f \equiv \gamma \ell^+ \ell^-$. Recall that a generic three-particle 
final state has no definite CP-parity. 
Nevertheless, the asymmetry (\ref{ACPtau-def}) which involves the integration over the three-particle phase space 
characterizes the CP violation as it
vanishes in a theory with a conserved CP. 

Taking into account meson oscillations, one obtains for time-dependent decay rates
\begin{eqnarray}
\label{tau-rate}
&&\frac{d\Gamma (B^0_q \to f)}{d\tau} = 
\frac{e^{-\,\Gamma\,\tau}}{2}\,
\Bigl [
A\, \textrm{ch}\left (y\,\Gamma\tau\right)\, +\,
B\, \cos\left (x\,\Gamma\tau\right)\, -\, 2\, C\, \textrm{sh}\left (y\,\Gamma\tau\right)\, -\,
    2\, D\, \sin\left (x\,\Gamma\tau\right)
\Bigr ], \\
&&\frac{d\Gamma (\bar B^0_q \to f)}{d\tau} = 
\frac{e^{-\,\Gamma\,\tau}}{2}\,
\Bigl [
A\, \textrm{ch}\left (y\,\Gamma\tau\right)\, -\,
B\, \cos\left (x\,\Gamma\tau\right)\,
-\, 2\, C\, \textrm{sh}\left (y\,\Gamma\tau\right)\, +\,
    2\, D\, \sin\left (x\,\Gamma\tau\right)
\Bigr ], \nonumber
 \end{eqnarray}
where the coefficients $A$, $B$, $C$, and $D$ may be expressed via the helicity amplitudes as follows: 
\begin{eqnarray}
\label{coeff-abcd}
&& A = \int d\hat s\,\tilde A(\hat s) =
\int\,\frac{d\Phi_3}{2\, M_1}
\sum_{\lambda_i\,\lambda_1\,\lambda_2}\,
\left (
\left |
 A^{(q)}_{\lambda_i\,\lambda_1\,\lambda_2}(\hat s,\,\cos\theta)
\right |^2 + 
\left | \bar A^{(q)}_{\lambda_i\,\lambda_1\,\lambda_2}(\hat s,\,\cos\theta)
\right |^2
\right), 
\\
&& B = \int d\hat s\,\tilde B(\hat s) =
\int\,\frac{d\Phi_3}{2\, M_1}
\sum_{\lambda_i\, \lambda_1\, \lambda_2}\,
\left (
\left |
A^{(q)}_{\lambda_i\,\lambda_1\,\lambda_2}(\hat s,\,\cos\theta)
\right |^2 - 
\left | \bar A^{(q)}_{\lambda_i\,\lambda_1\,\lambda_2}(\hat s,\,\cos\theta)
\right |^2
\right),
\nonumber \\
&& C = \int d\hat s\,\tilde C(\hat s) =
\int\,\frac{d\Phi_3}{2\, M_1}
\sum_{\lambda_i\, \lambda_1\, \lambda_2}\, 
\textrm{Re}\,\left (e^{-2i\phi_{\rm ckm}}
     A^{(q)\, *}_{\lambda_i\,\lambda_1\,\lambda_2}(\hat s,\,\cos\theta)
     \bar A^{(q)}_{\lambda_i\,\lambda_1\,\lambda_2}(\hat s,\,\cos\theta)
    \right),
\nonumber \\
&& D = \int d\hat s\,\tilde D(\hat s) =
\int\,\frac{d\Phi_3}{2\, M_1}
\sum_{\lambda_i\, \lambda_1\, \lambda_2}\,
\textrm{Im}\,\left (e^{-2i\phi_{\rm ckm}}
     A^{(q)\, *}_{\lambda_i\,\lambda_1\,\lambda_2}(\hat s,\,\cos\theta)
    \bar A^{(q)}_{\lambda_i\,\lambda_1\,\lambda_2}(\hat s,\,\cos\theta)
    \right).  \nonumber 
\end{eqnarray}
In the expressions above one should use the appropriate helicity amplitudes and phase-space factors: 
for the case of semileptonic decays they are given in Appendix B; for radiative leptonic decays -- in Appendix C. 
As already noticed above the helicity amplitudes of Appendix C contain the contributions of the 
Bremsstrahlung diagrams. 

The weak phase $\phi_{\rm ckm}$ is related to the CP-violating phase of the CKM matrix:
$V^*_{tb}V_{tq}=|V^*_{tb}V_{tq}|e^{-i\phi_{\rm ckm}}$. 
For instance, for the case $q=d$, $\phi_{\rm ckm}=\beta$ ($\equiv \phi_1$ of Belle) \cite{stone}.

We now take into account the existence of two mass-eigenstates of neutral $B$-mesons. 
These states have the masses $M_{\ell}$ and $M_h$, and the full widths  $\Gamma_{\ell}$ and $\Gamma_h$, respectively.
We use the definitions
$$
M=(M_{\ell}\, +\, M_h)/2,\ \Delta m = M_h - M_{\ell},\
\Gamma = (\Gamma_{\ell} + \Gamma_h)/2,\ \Delta\Gamma = \Gamma_{\ell} - \Gamma_h, \
x = \Delta m/\Gamma,\ y = \Delta\Gamma/\Gamma.
$$ 
Then, we obtain the following expression for time-dependent asymmetry:
\begin{eqnarray}
\label{ACPtau-equation}
A^{B_q\to f}_{CP}(\tau) &=&
\frac{2\, D\, \sin\left (x\,\Gamma\tau\right)\, -\, 
          B\, \cos\left (x\,\Gamma\tau\right)}
     {    A\, \textrm{ch}\left (y\,\Gamma\tau\right)\, -\,
      2\, C\, \textrm{sh}\left (y\,\Gamma\tau\right)}\, \equiv \frac{S_f \sin \left (\Delta m\,\tau \right )\, -\, 
      C_f \cos \left (\Delta m\,\tau \right )}
         {\textrm{ch}\left (\Delta\Gamma\,\tau\right)\, -\, 
      H_f \textrm{sh}\left (\Delta\Gamma\,\tau\right)}.
\end{eqnarray}
The coeffcients $S_f$, $C_f$, and $H_f$, which are used for the parameterization of the data,
may be expressed via $A$, $B$, $C$, and $D$. 

Let us make the following remarks: 

(i) For $B_d$-mesons $\Delta\Gamma \ll \Gamma$ and the denominator (\ref{ACPtau-equation}) is almost equal to one. 
For $B_s$-mesons the contribution of $H_f$ is essential \cite{pdg2008}.

(ii) A nonzero asymmetry arises from $\bar B^0_q\, B^0_q$--oscillations, from the contributions of 
$u\bar u$-, $c\bar c$-pairs, $\rho$, $\omega$, $\phi$, and $c\bar c$ vector resonances ($J/\psi$, 
$\psi'$, etc), and the weak annihilation. Since the oscillation frequency for $B_s$-mesons is much larger than that for 
$B_d$-mesons, and the weak-annihilation in $B_s$-decays is negligible, one finds $A^{B_s}_{CP}(\tau)\ll A^{B_d}_{CP}(\tau)$. 

Time-independent $CP$-asymmetry may be represented via $\tilde A(\hat s)$, ..., $\tilde D(\hat s)$ 
from Eq.~(\ref{coeff-abcd}) as follows: 
\begin{eqnarray}
\label{ACPs-equation}
A^{B_q \to f}_{CP}(\hat s)\, =\, 
\frac{\displaystyle{\frac{d\Gamma(\bar B_q\,\to\, f)}{d\hat s}} - \frac{d\Gamma(B_q\,\to\, f)}{d\hat s}}
     {\displaystyle{\frac{d\Gamma(\bar B_q\,\to\, f)}{d\hat s}} + \frac{d\Gamma(B_q\,\to\, f)}{d\hat s}}\, 
=\, -\,
\left(\frac{1 - y^2}{1 + x^2}\right)
\frac{\tilde B(\hat s) - 2 x \tilde D(\hat s)}{\tilde A(\hat s) - 2 y \tilde C(\hat s)}. 
\end{eqnarray}

\section{Numerical results\label{Sect:5}}
We are going to apply now the formulas derived above and to provide numerical results for the asymmetries. 
We use the following numerical parameters: 

(i) Table \ref{table:Bparam} summarizes the parameters of the $B^0_{d, s}$-oscillations which we use for our numerical estimates. 
\begin{table}[tbh]
\caption{Parameters of $B^0_{d,\, s}$-oscillations  \cite{pdg2008,mxz2008,stone}
\label{table:Bparam}}
\begin{tabular}{|l|c|c|}
\hline 
$B$-meson parameters                 & $B^0_d$        & $B^0_s$  \\
\hline 
B-meson mass $M_1$ (GeV)                   & $5.28$         & $5.37$   \\
\hline 
Width $\Gamma$ (ps$^{-1}$)                 & $0.65$         & $0.67  $ \\
\hline 
Mass difference $\Delta m$ (ps$^{-1}$)       & $0.507$        & $17.77$   \\
\hline 
Width difference $\Delta\Gamma$ (ps$^{-1}$)  & $0.005  $      & $0.1$    \\
\hline
\end{tabular}
\end{table}

(ii) The Wilson coefficients for the SM are evaluated at $\mu=5$ GeV \cite{wc}
for $C_2(M_W) = -1$:  $C_1(\mu)=\, 0.241$, $C_2(\mu)=\, -1.1$,
$a_1(\mu) =\, -0.126$, $C_{7\gamma}(\mu)=\, 0.312$, $C_{9V}(\mu)=\, -4.21$ 
and $C_{10A}(\mu)=4.64$. Respectively, we use the running quark masses in the $\overline{\rm MS}$ 
scheme at the same scale: $m_b=4.2$ GeV, $m_s=60-80$ MeV, and the $d$-quark mass is neglected. 
For the coefficient $C_{eV}^{\rm eff}$ we use the model proposed in \cite{krugerCP} which 
takes into account resonances in a gauge-invariant way. 

For the CKM matirx elements we use the values reported in the 2008 edition of PDG \cite{pdg2008}: 
$A=0.814$, $\lambda=0.226$, $\bar \rho=0.135$, $\bar \eta=0.35$. 

(iii) We make use of the form factor parameterizations for rare semileptonic decays from \cite{ms}
and for rare radiative decays from \cite{mk,mnt}. The accuracy of these predictions for the form factors 
is expected to be at the level of 10-15\%, which influences strongly the predictions for the decay rates. 
However, the form factor uncertainties cancel to a large extent in the asymmetries which therefore can be 
predicted with a few percent accuracy \cite{mns,mns2}. 
For the decay constants of the $B$-mesons we use the values   
$f_B= 220\pm 20$ MeV and $f_{B_s}= 240\pm 20$ MeV. 

(iv) A cut on the Bremsstrahlung photon spectrum at 20 MeV in the $B$-meson rest frame is applied. 
This corresponds to the expected level of the photon energy resolution of the LHCb detector. 

\subsection{Forward-backward asymmetry}
\begin{figure}[!t]
\begin{center}
\begin{tabular}{cc}
\includegraphics[width=5.8cm]{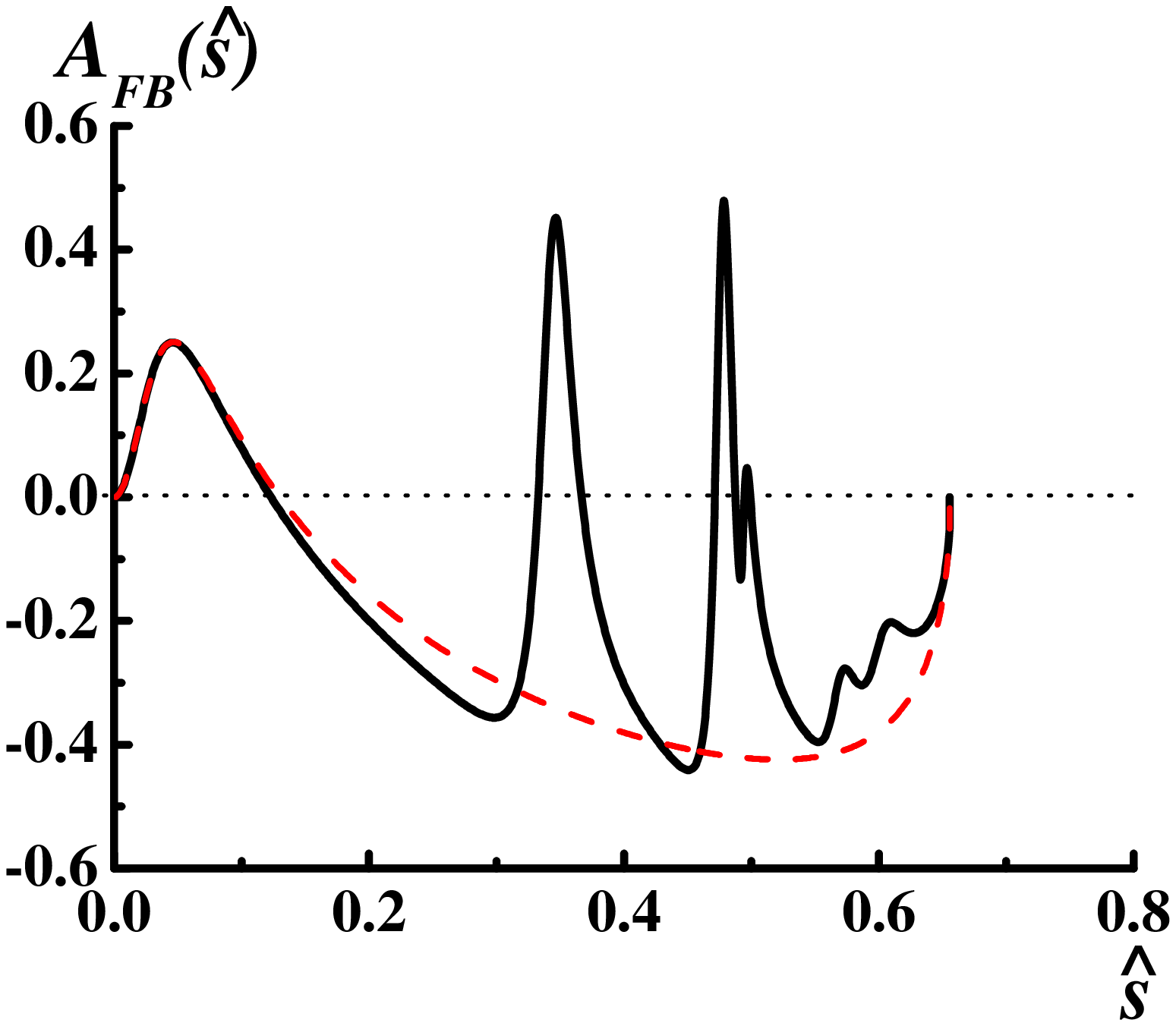} 
& 
\includegraphics[width=5.8cm]{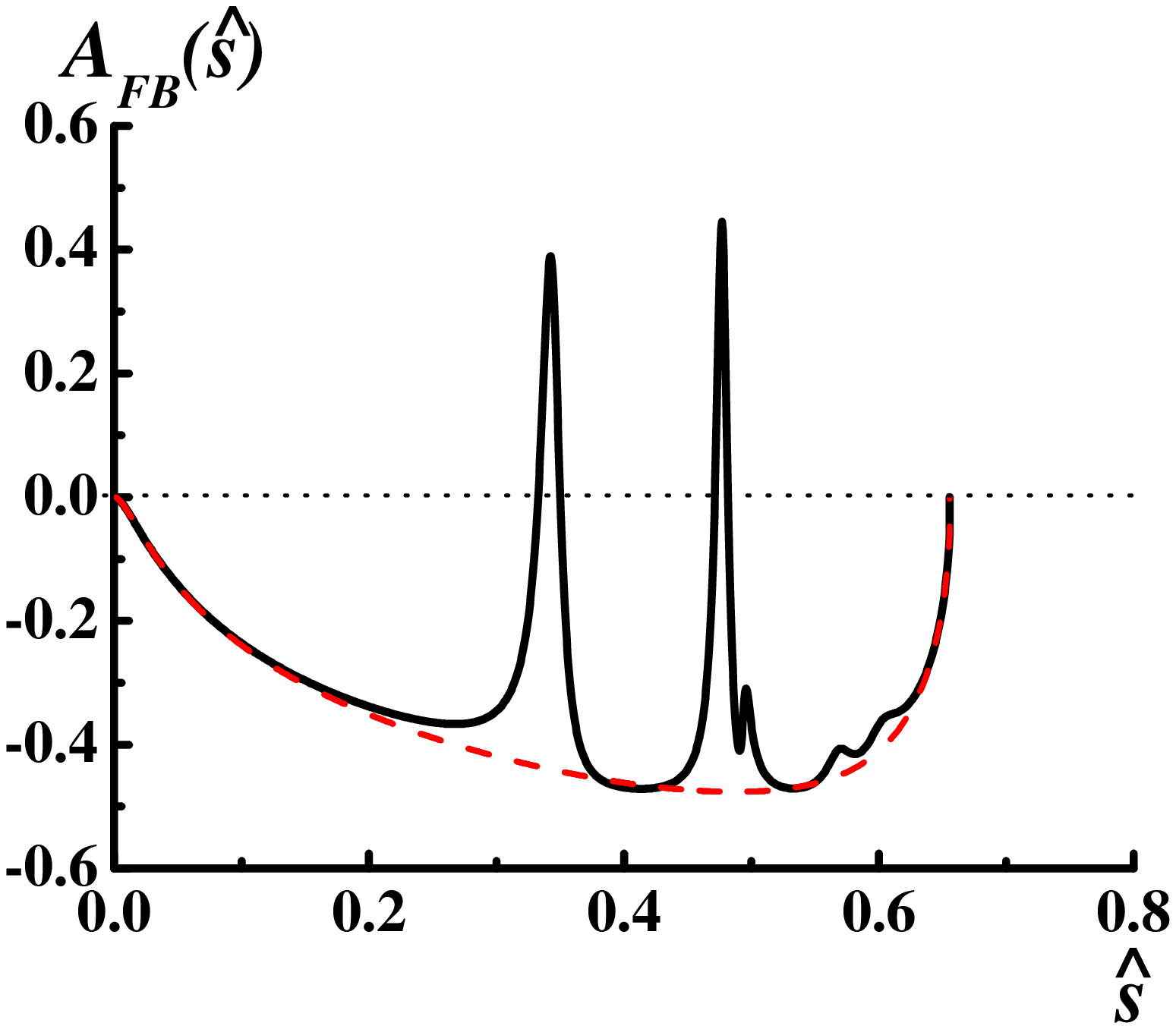} 
\\
(a) & (b)
\\
\includegraphics[width=5.8cm]{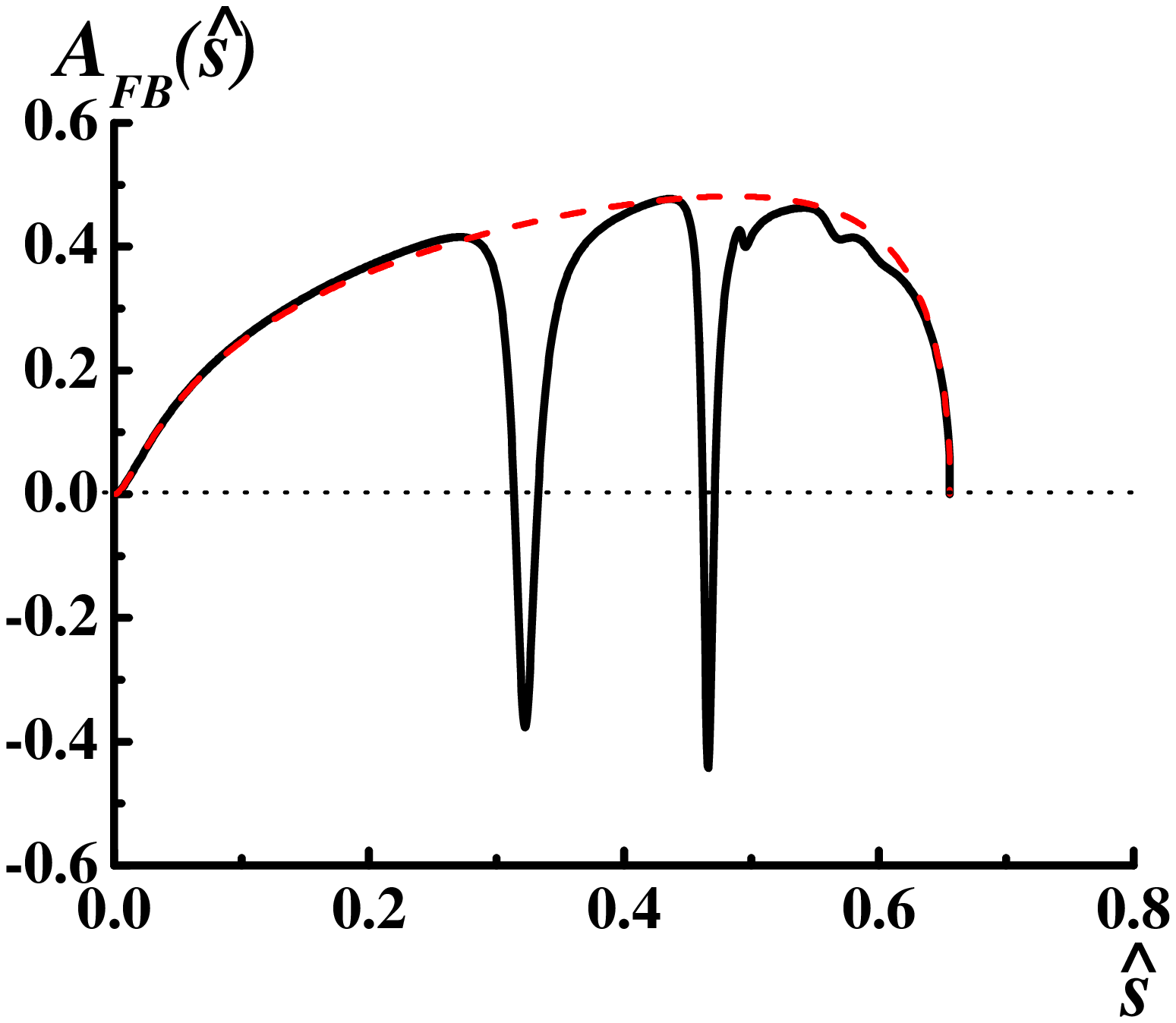} 
& 
\includegraphics[width=5.8cm]{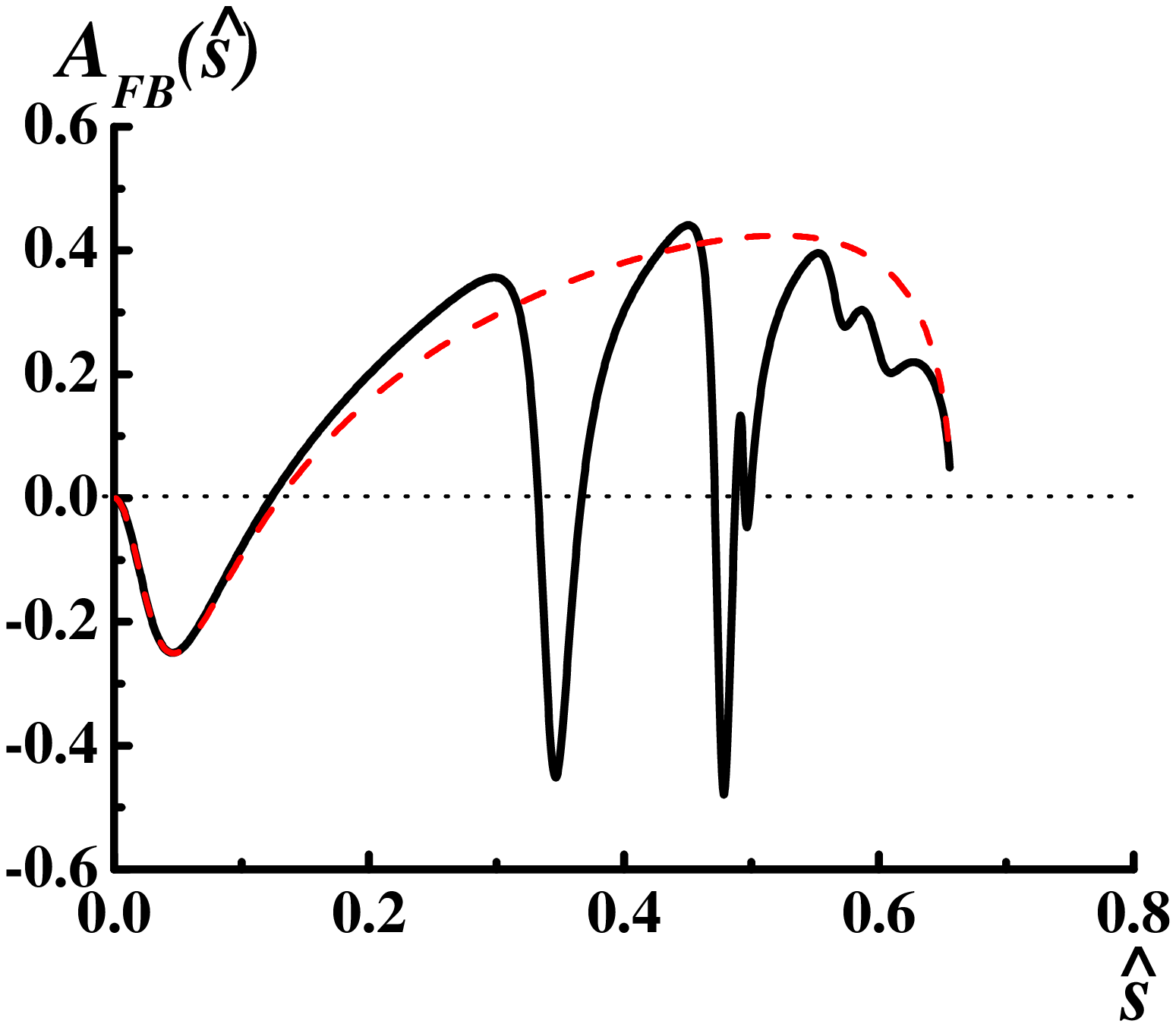}
\\
(c) & (d)
\end{tabular}
\end{center}
\caption{\label{Fig:1}
$A_{FB}$ for rare semileptonic $\bar B_s\to \phi\mu^+\mu^-$ decays: (a) in the SM;  
(b) For $C_{7\gamma}=-C^{\rm SM}_{7\gamma}$, (c) For $C_{9V}=-C^{\rm SM}_{9V}$, (d) For $C_{10A}=-C^{\rm SM}_{10A}$. 
Solid line (black): the full asymmetry which takes into account the $J/\psi$, $\psi'$, etc contributions. 
Dashed line (red): the non-resonant asymmetry.}
\end{figure}

\begin{figure}[!h]
\begin{center}
\begin{tabular}{cc}
\includegraphics[width=5.8cm]{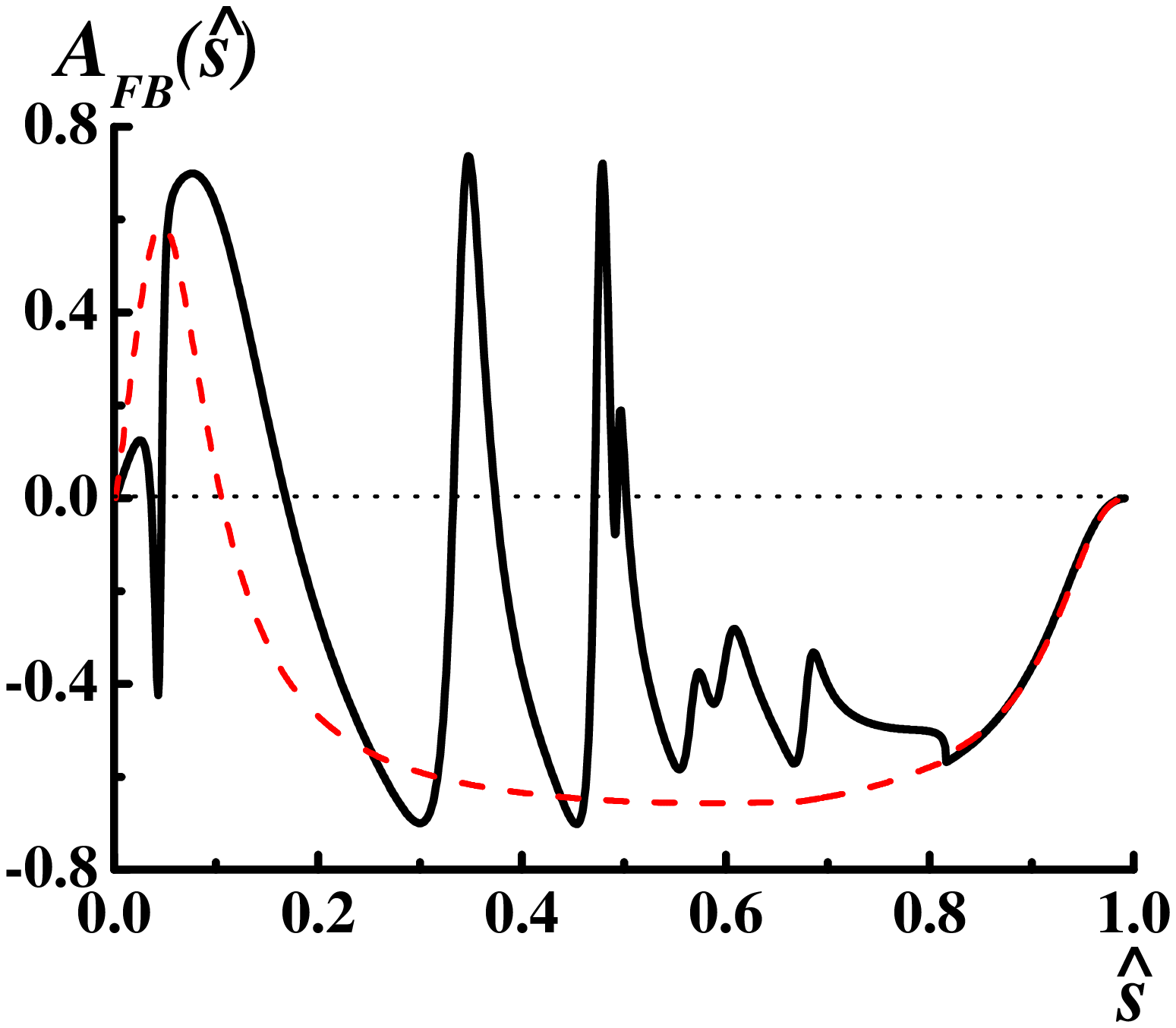} 
& 
\includegraphics[width=5.8cm]{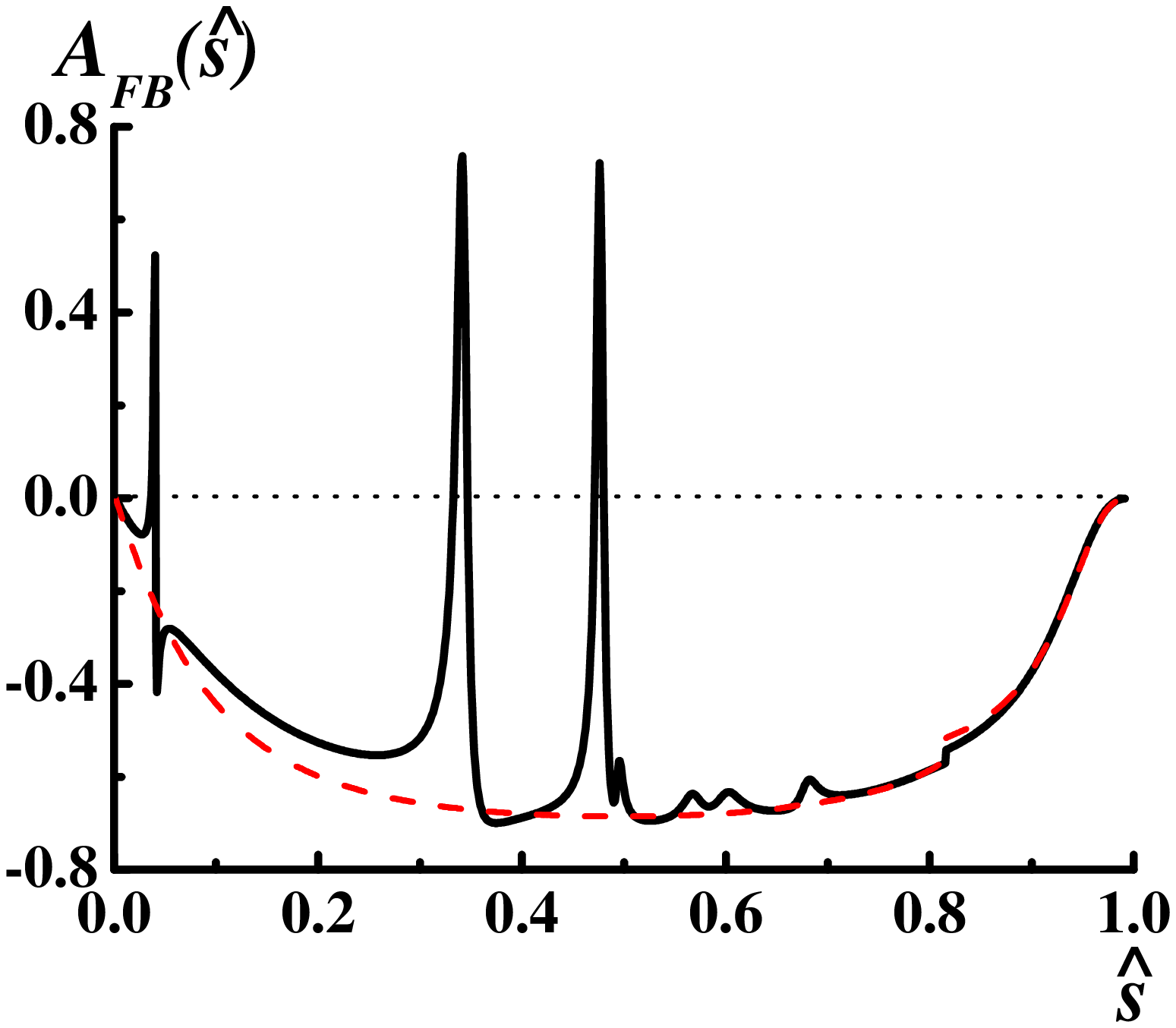} 
\\
(a) & (b)
\\
\includegraphics[width=5.8cm]{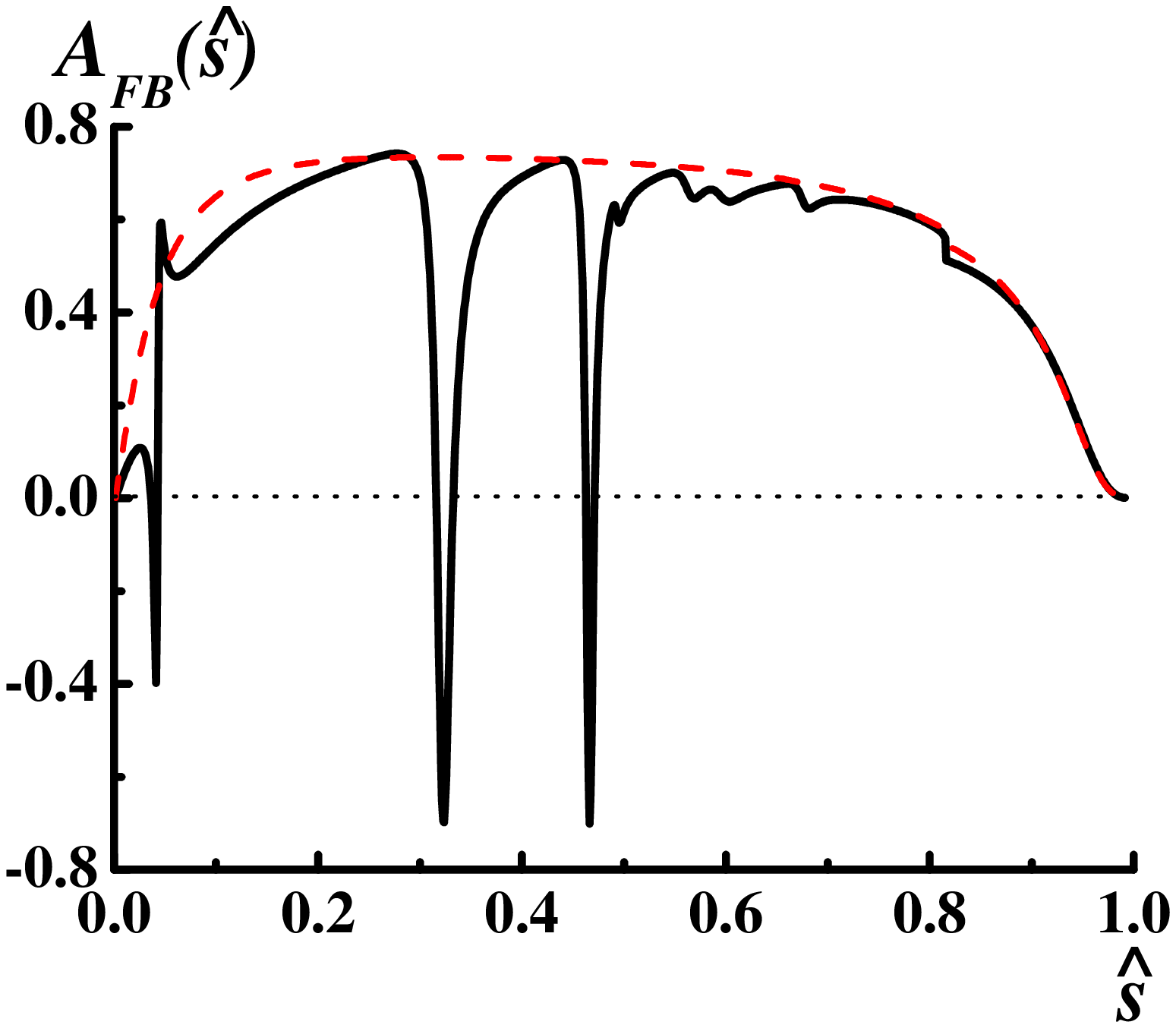} 
& 
\includegraphics[width=5.8cm]{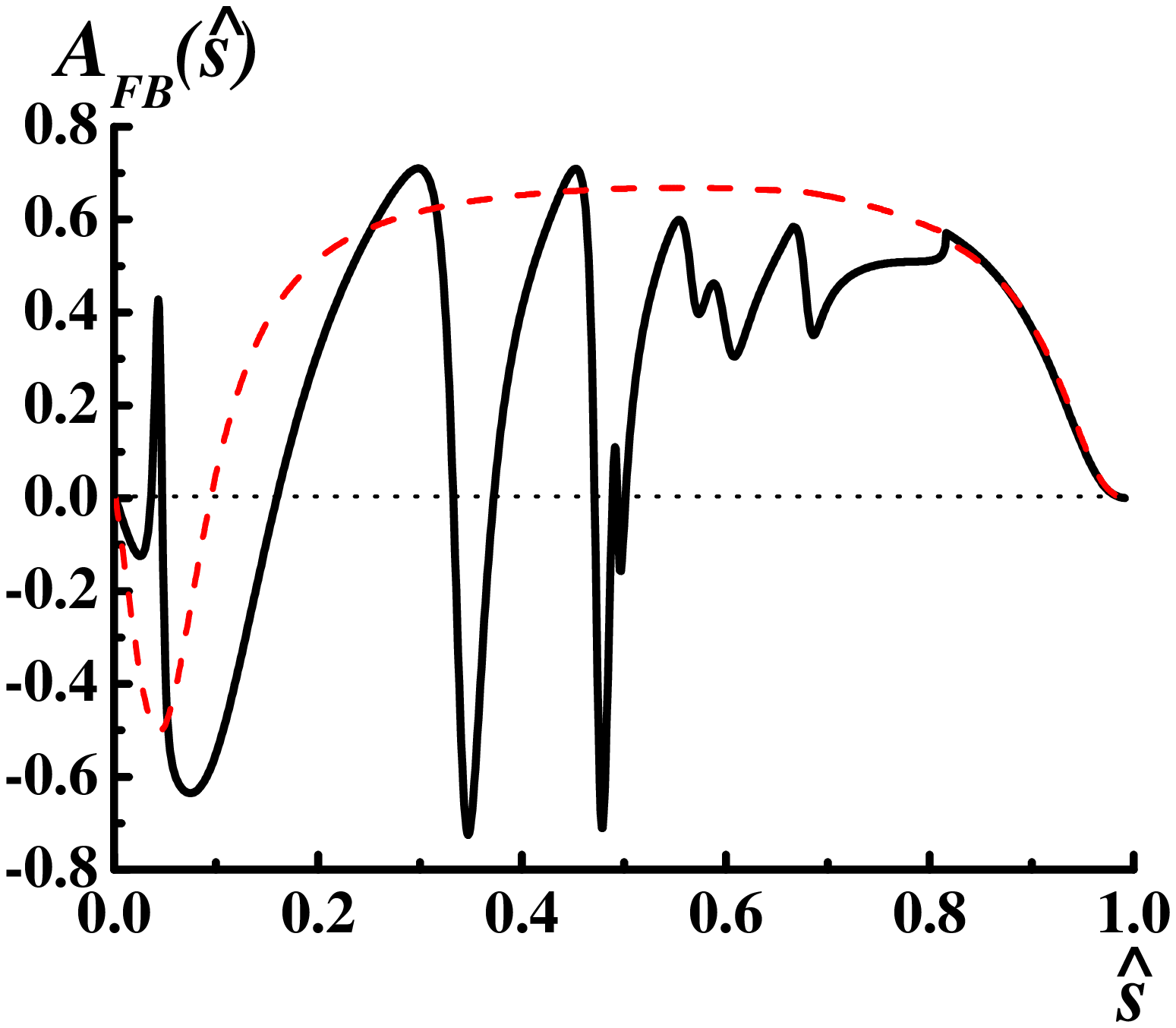}
\\
(c) & (d)
\end{tabular}
\end{center}
\caption{\label{Fig:2}
$A_{FB}(\bar s)$ for $\bar B_s\to \gamma\mu^+\mu^-$ decays: 
(a) In the SM.   
(b) For $C_{7\gamma}=-C^{\rm SM}_{7\gamma}$. 
(c) For $C_{9V}=-C^{\rm SM}_{9V}$. 
(d) For $C_{10A}=-C^{\rm SM}_{10A}$. 
Solid line (black): the asymmetry calculated for the full amplitude of Ref.~\cite{mnt}. 
Dashed line (red): the asymmetry calculated for the amplitude without the contributions of light neutral vector mesons 
$\phi$, the $c\bar c$ resonances ($J/\psi$, $\psi'$, \ldots), Bremsstrahlung, and the weak annihilation.}
\end{figure}

\begin{figure}
\begin{center}
\begin{tabular}{cc}
\includegraphics[width=5.8cm]{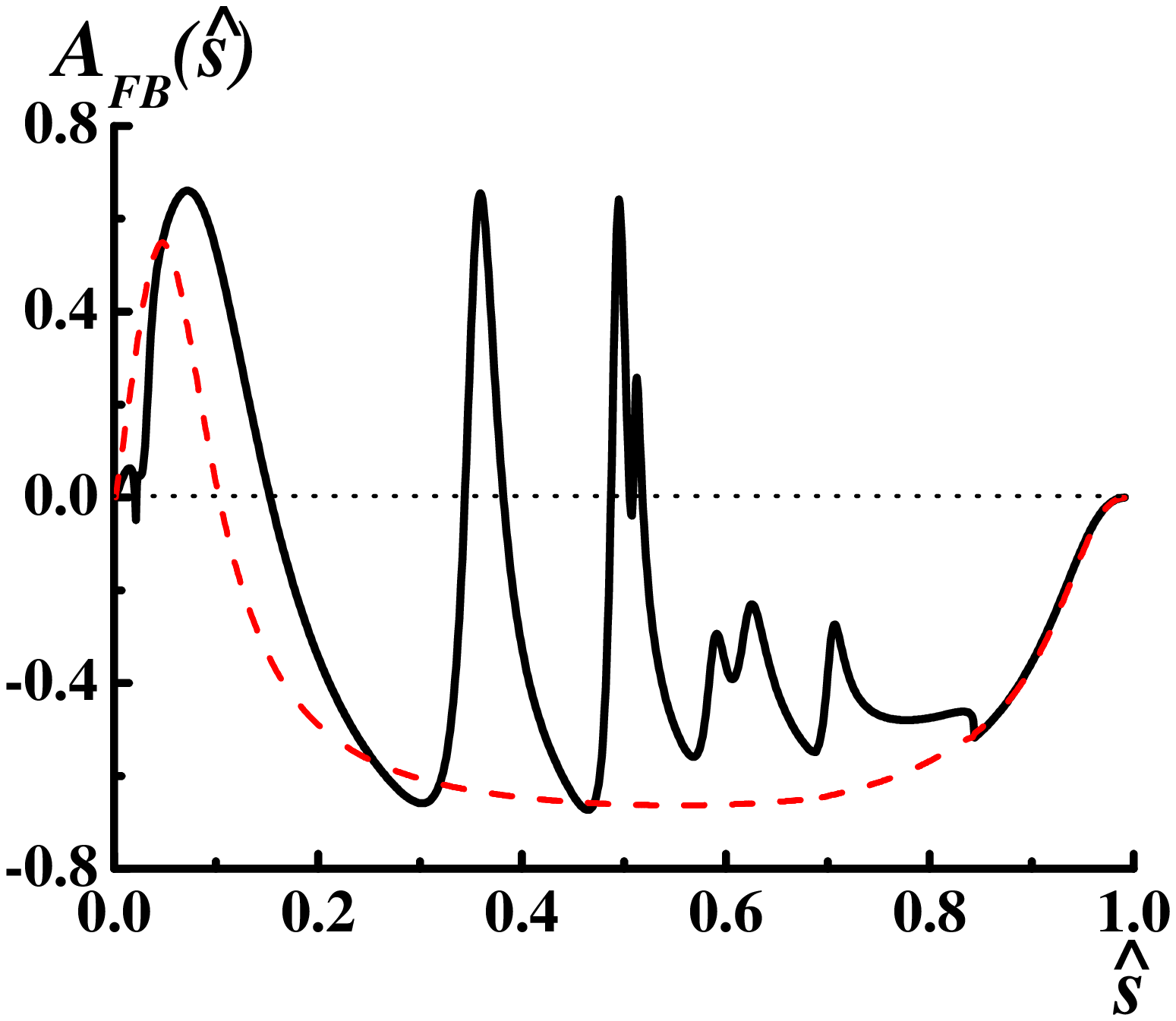} 
& 
\includegraphics[width=5.8cm]{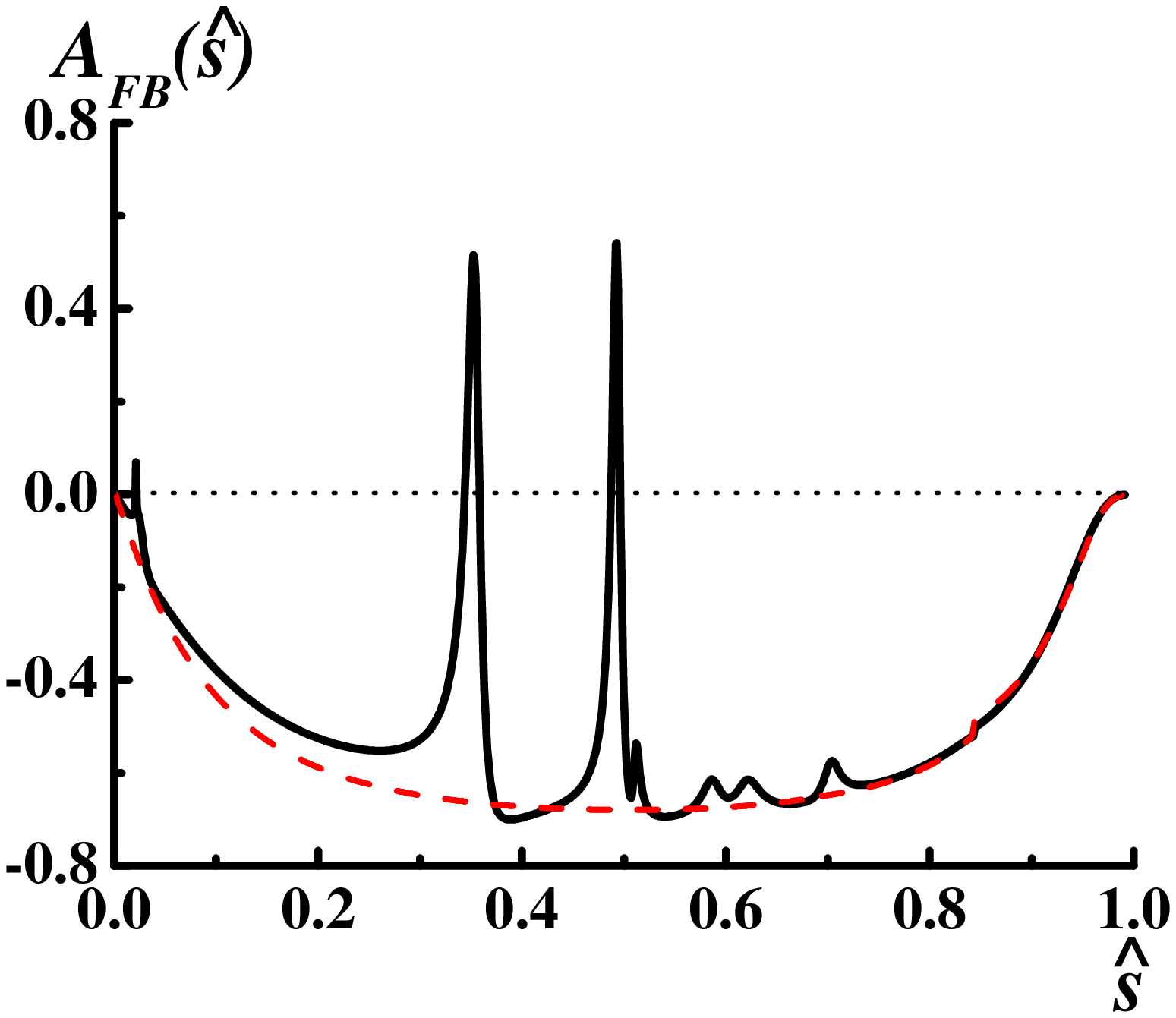} 
\\
(a) & (b)
\\
\includegraphics[width=5.8cm]{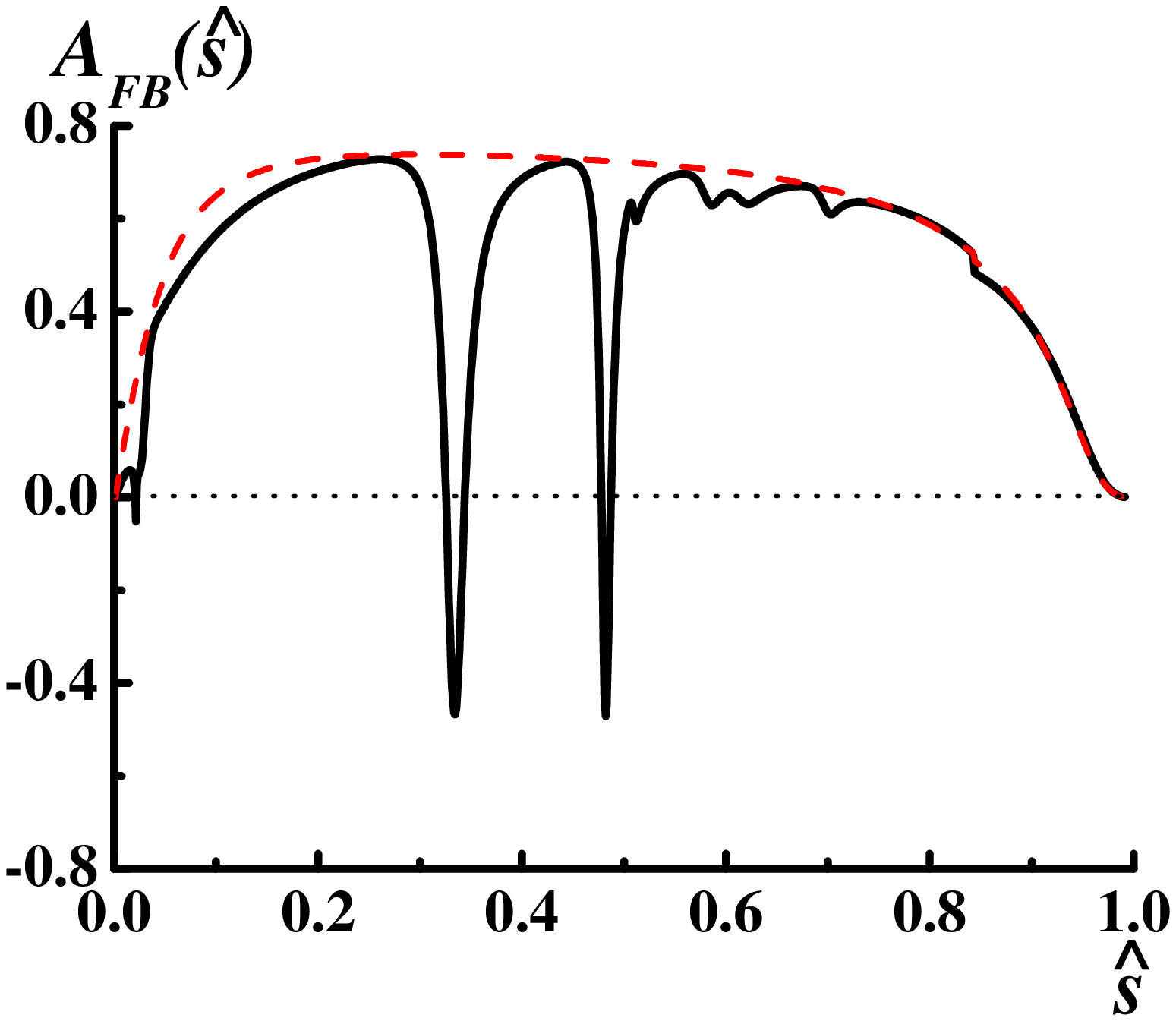} 
& 
\includegraphics[width=5.8cm]{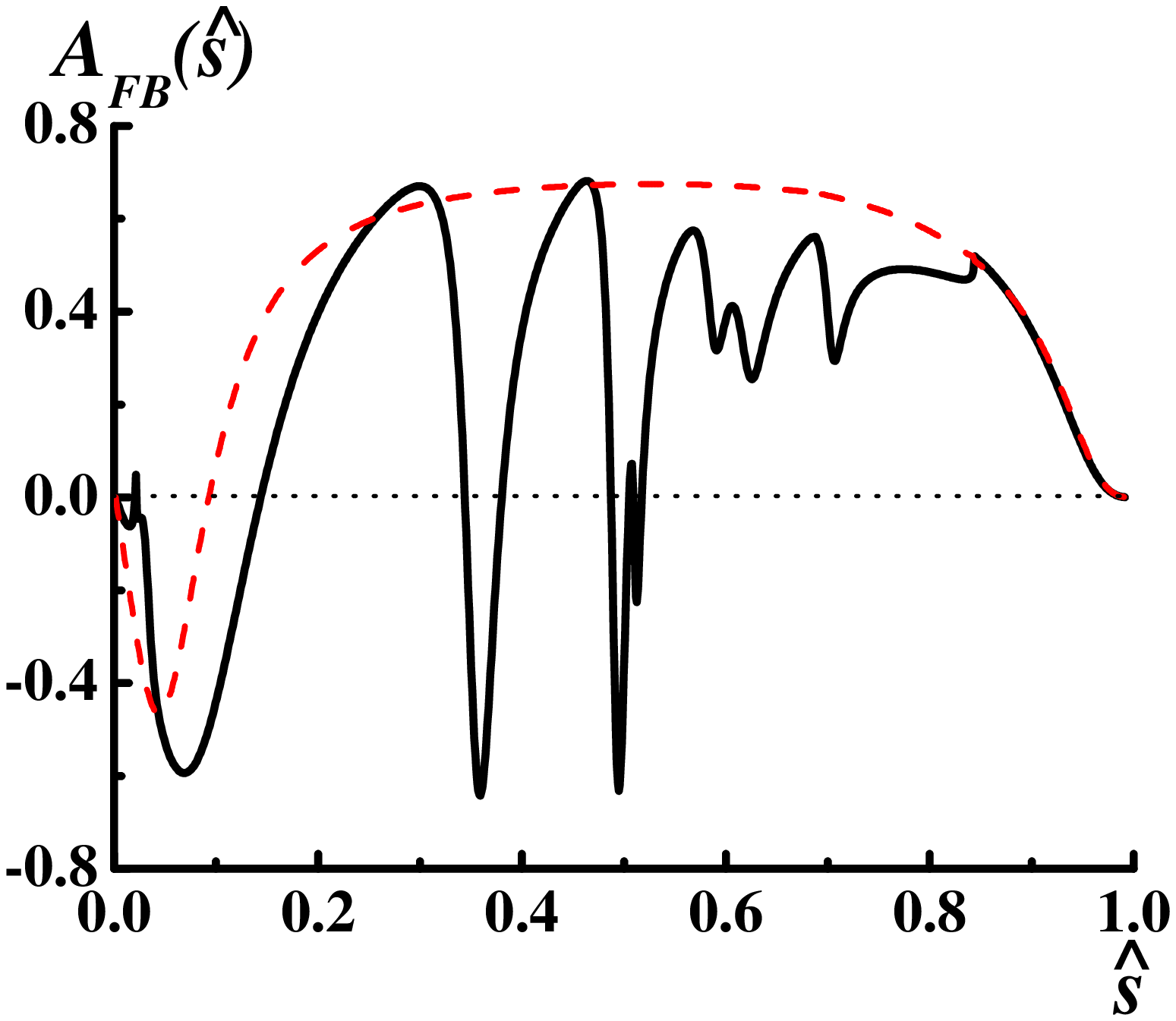}
\\
(c) & (d)
\end{tabular}
\end{center}
\caption{ \label{Fig:3}
$A_{FB}(\bar s)$ for $\bar B_d\to \gamma\mu^+\mu^-$ decays: 
(a) In the SM.  
(b) For $C_{7\gamma}=-C^{\rm SM}_{7\gamma}$. 
(c) For $C_{9V}=-C^{\rm SM}_{9V}$. 
(d) For $C_{10A}=-C^{\rm SM}_{10A}$. 
Solid line (black): the asymmetry calculated for the full amplitude of Ref.~\cite{mnt}. 
Dashed line (red): the asymmetry calculated for the amplitude without the contributions of light neutral vector mesons 
$\omega$, $\rho^0$, the $c\bar c$ resonances ($J/\psi$, $\psi'$, \ldots) Bremsstrahlung, and the weak annihilation.}
\end{figure}

The calculated forward-backward asymmetries are presented in Figs.~\ref{Fig:1}--\ref{Fig:3}. 

The decay $\bar B_s\to \phi\mu^+\mu^-$ (Fig.~\ref{Fig:1}) is of special interest: 
the detector LHCb may accumulate sufficient 
data sample for this decay already after the first few months of operation. 
Qualitatively, the asymmetry has the same structure as in the 
$B\to K^*\mu^+\mu^-$ decays: its behaviour at small $\hat s$ 
is sensitive to the invertion of the signs of $C_{7\gamma}$ and $C_{10A}$ compared to the SM. 
Fig.~\ref{Fig:1} shows the influence of the sign invertion in one of the Wislon coefficients compared to the SM. 
      
Figs.~\ref{Fig:2} and \ref{Fig:3} present $A_{FB}$ for the radiative decays $\bar B_s\to \gamma\mu^+\mu^-$ and 
$\bar B_d\to \gamma\mu^+\mu^-$, respectively. 
Qualitatively, the asymmetry behaves at large and intermediate $\hat s$ similarly to the $B_s\to \phi\mu^+\mu^-$ decay, 
although has a larger magnitude. At small $\hat s$, however, the assymetry is influenced by the 
neutral light vector resonances $\phi$, $\omega$, and $\rho^0$ \cite{mnt}: 
these resonances cause a strong distortion of the full asymmetry compared to a nonresonance asymmetry. 
In particular, they leads to a visible shift of 
the ``zero-point'' compared to its location in the non-resonant asymmetry, which may 
be reliably calculated in the SM \cite{null-test}.     

\newpage                       

\newpage

\subsection{CP-violating asymmetries}
We present now the time-independent and the time-dependent CP-asymmetries in 
$B_d\to (\rho,\gamma) \mu^+\mu^-$. Concerning the $B_s\to (\phi,\gamma) \mu^+\mu^-$ decays we would 
like to mention the following: we have calculated these asymmetries and found that  
$A_{CP}(\hat s)$, mainly due to flavor oscillations of the $B_s$ mesons, 
is extremely small (smaller than 0.1\%) and therefore cannot be studied experimentally;  
$A_{CP}(\tau)$ is not small but measuring this asummetry would require time resolution much smaller 
than the $B_s$ lifetime. 

%
\subsubsection{Time-independent asymmetry}
\begin{figure}[!ht]
\begin{center}
\begin{tabular}{cc}
\includegraphics[width=8.2cm]{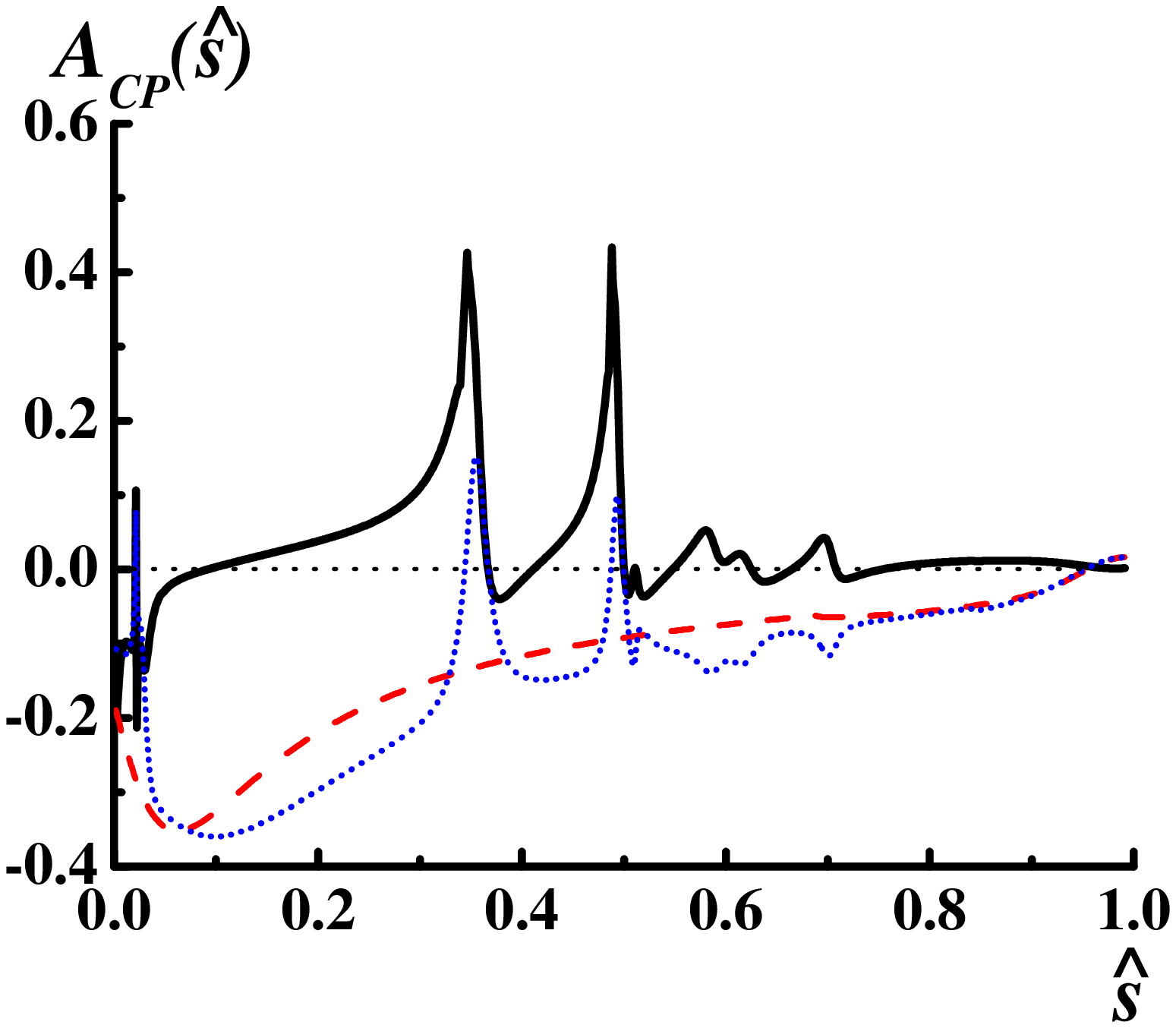} & 
\includegraphics[width=8.2cm]{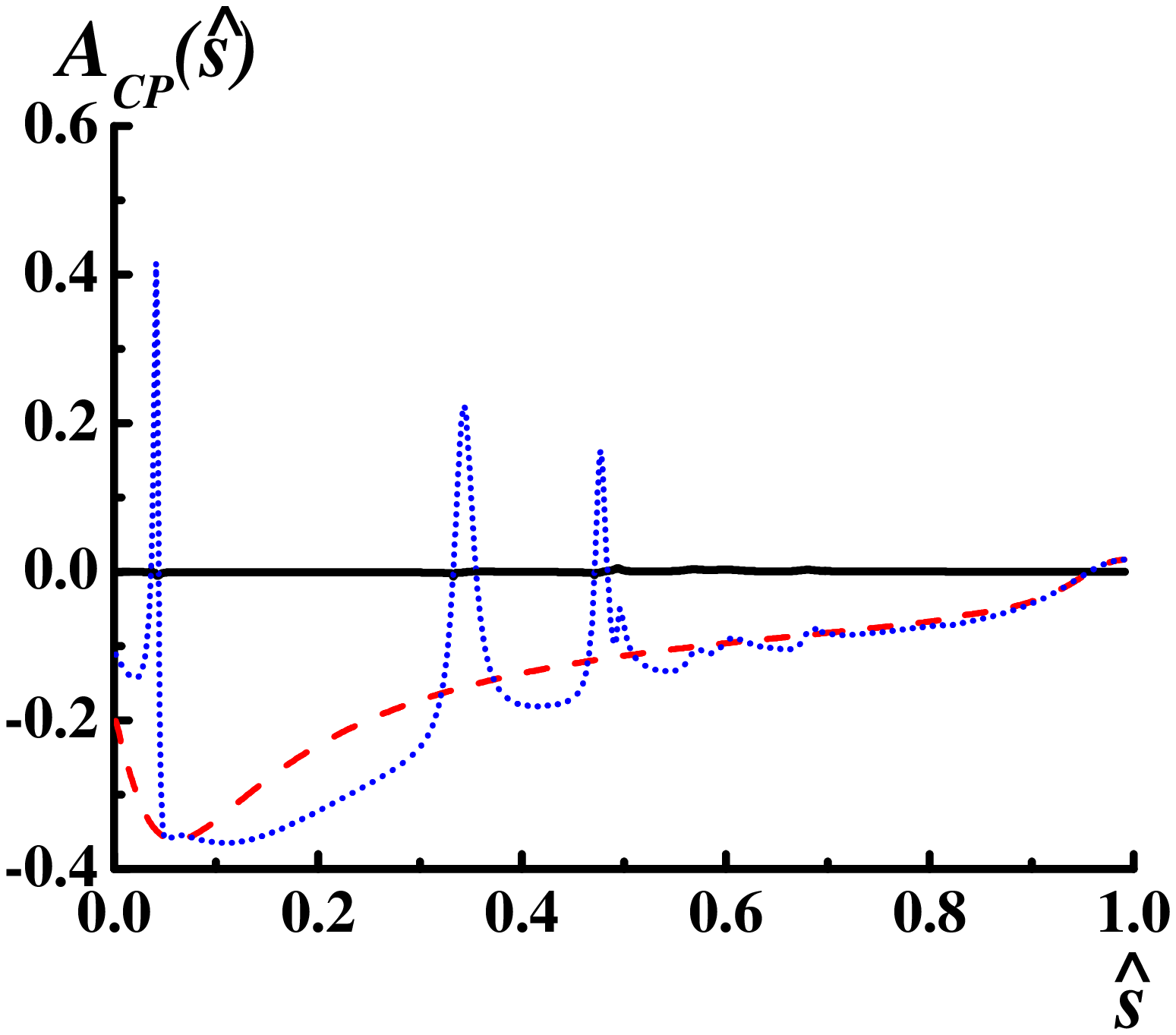} 
\\
(a) & (b)
\end{tabular}
\end{center}
\caption{\label{Fig:4}
The influence of $B$-meson flavor oscillations upon CP-violating asymmetries: 
(a) $B_d\to \gamma\mu^+\mu^-$, 
(b) $B_s\to \gamma\mu^+\mu^-$.
Dashed line (red): $A_{CP}$ without resonances and without flavor oscillations; 
Dotted line (blue): $A_{CP}$ with resonances but without flavor oscillations; 
Solid line (black): $A_{CP}$ after flavor oscillations have been taken into account.}
\end{figure}

First, we would like to demonstrate the impact of flavor oscillations of the initial mesons on the 
resulting CP-violating asymmetries.
Fig.~\ref{Fig:4} shows $A_{CP}(\hat s)$ for $B_{d,s}\to\gamma\mu^+\mu^-$ decays. Obviously, flavour oscillations 
lead to a strong suppression of the the resulting CP-violating asymmetries in $B_d$ decays and to a complete
vanishing of $A_{CP}$ in $B_s$ decays.  

\begin{figure}[!ht]
\begin{center}
\begin{tabular}{ccc}
\includegraphics[width=5.8cm]{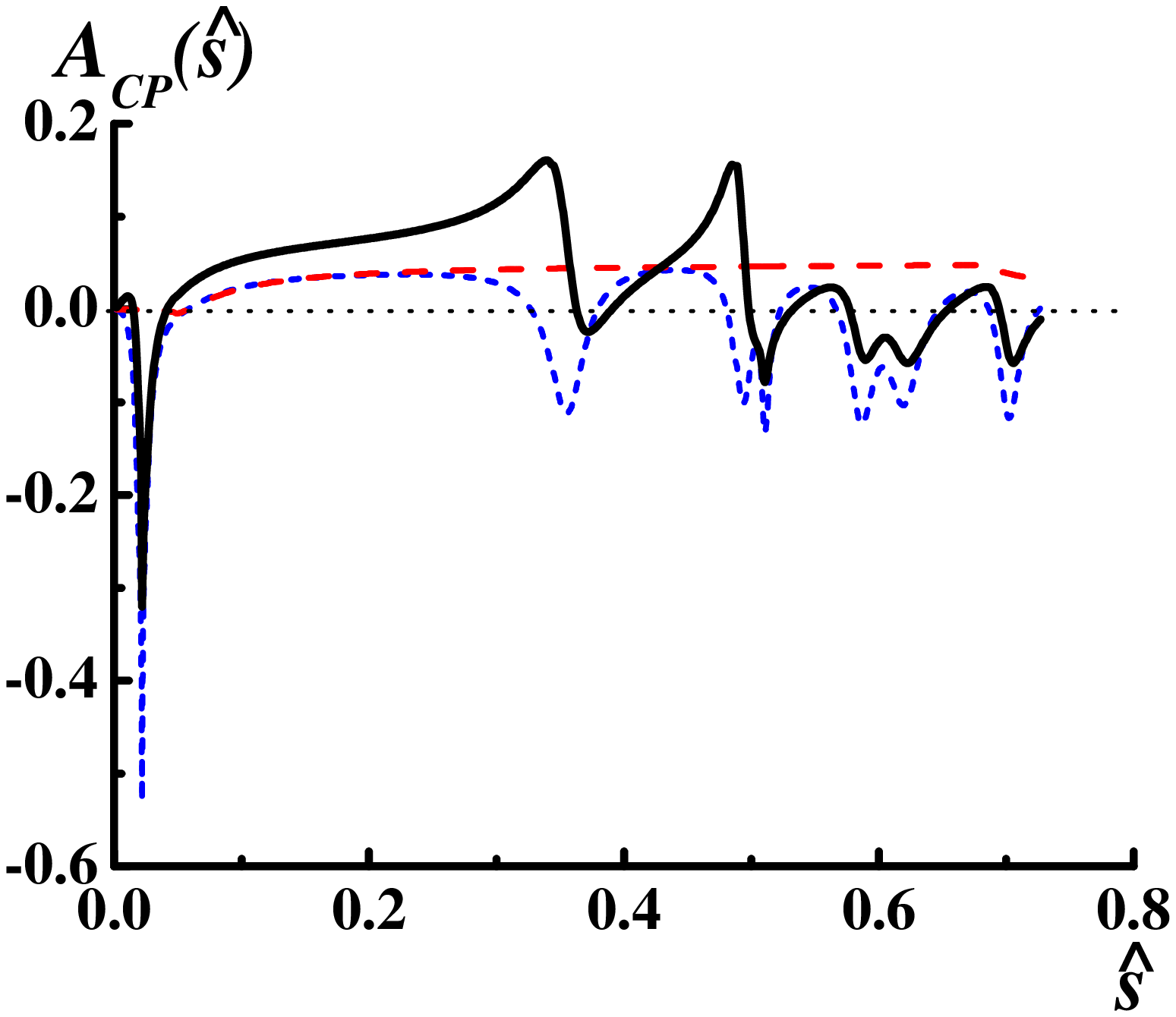} & 
\includegraphics[width=5.8cm]{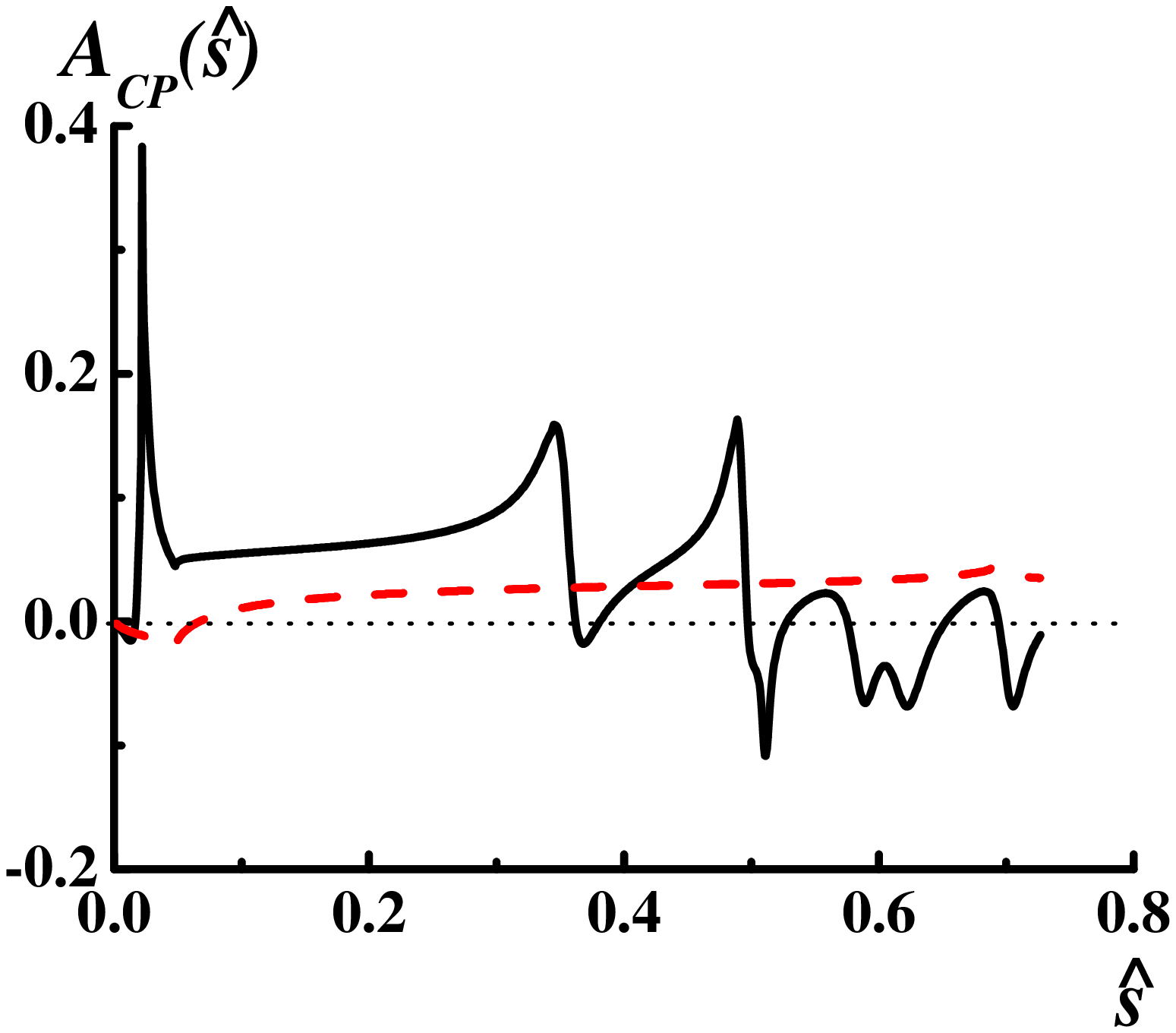} &
\includegraphics[width=5.8cm]{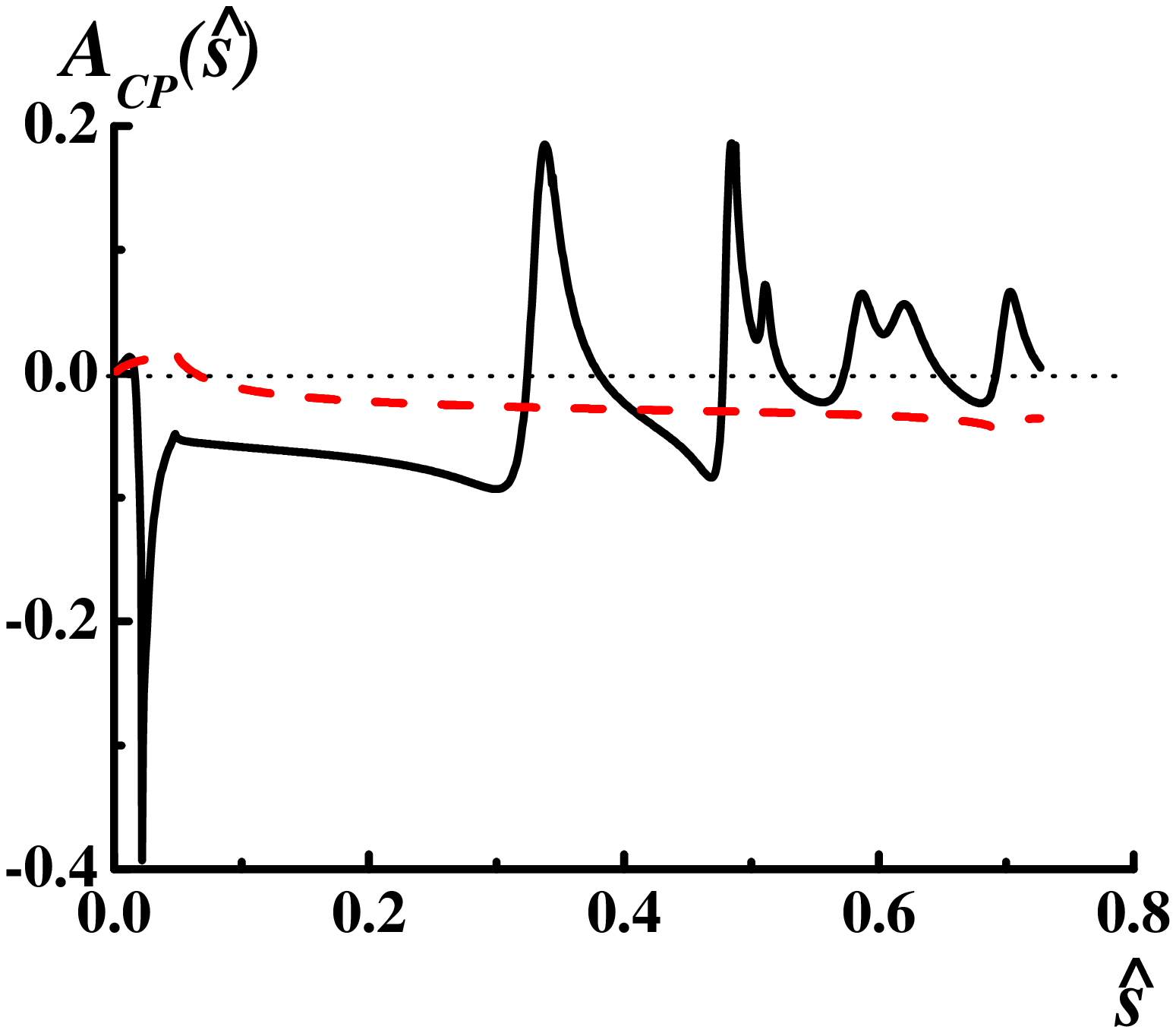} 
\\
(a) & (b) & (c)
\end{tabular}
\end{center}
\caption{\label{Fig:4.1}
Time-independent CP-asymmetry $A_{CP}(\hat s)$ in $B_d\to \rho\mu^+\mu^-$ decays.
(a) SM  (b) $C_{7\gamma}=-C^{\rm SM}_{7\gamma}$ (c) $C_{9V}=-C^{\rm SM}_{9V}$. Flavor oscillations have been taken into account. 
Solid line (black) line: full asymmetry. 
Dashed (red) line: nonresonant asymmetry.
Dotted (blue) line shows the asymmetry if flavor oscillations are not taken into account.}
\end{figure}

\begin{figure}[!hb]
\begin{center}
\begin{tabular}{ccc}
\includegraphics[width=5.8cm]{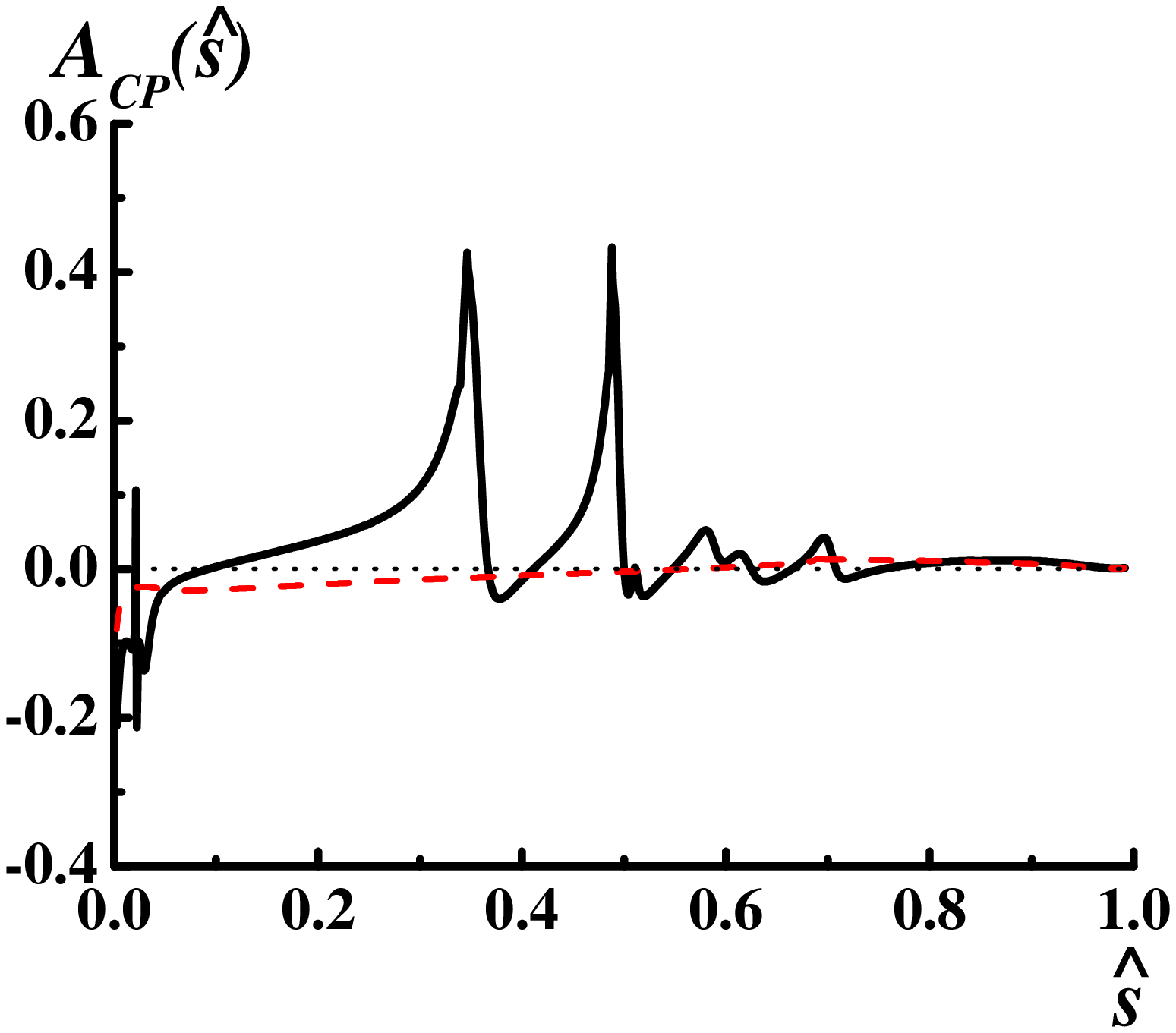} & 
\includegraphics[width=5.8cm]{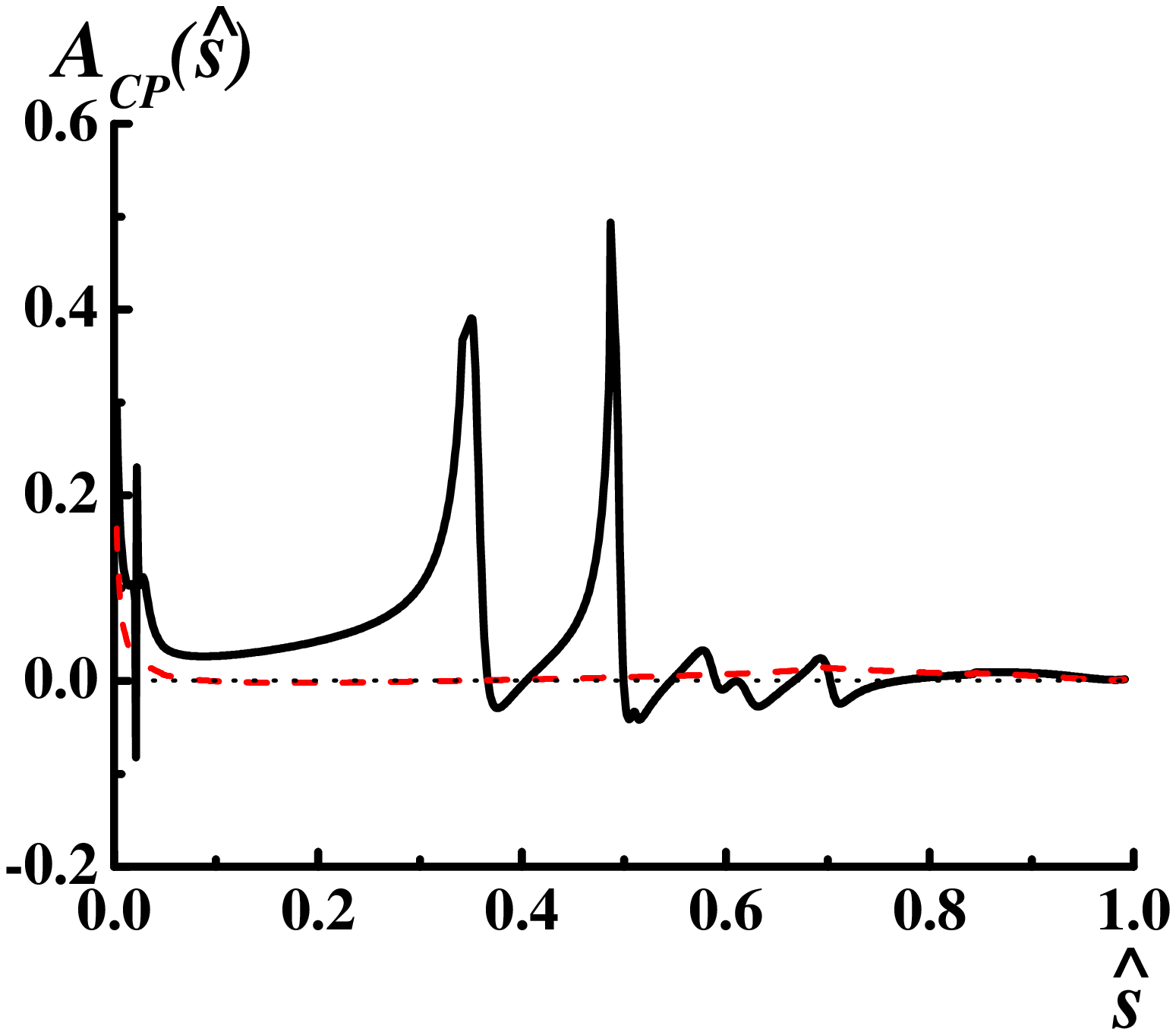} &
\includegraphics[width=5.8cm]{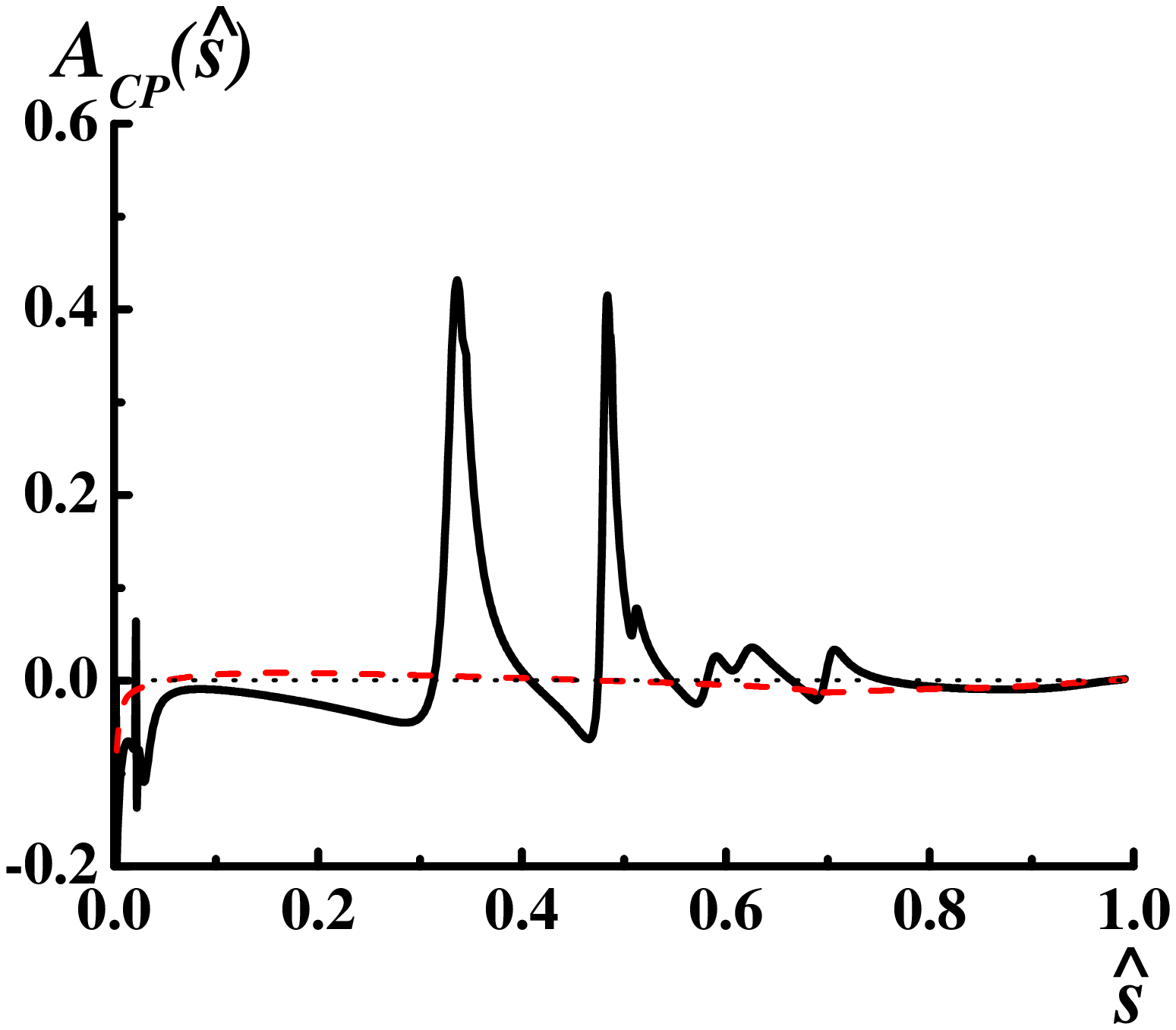}
\\
(a) & (b)& (c)
\\
\end{tabular}
\end{center}
\caption{\label{Fig:4.2}
Time-independent CP-asymmetry $A_{CP}(\hat s)$ in $B_d\to \gamma\mu^+\mu^-$ decays.
(a) SM  (b) $C_{7\gamma}=-C^{\rm SM}_{7\gamma}$ (c) $C_{9V}=-C^{\rm SM}_{9V}$. 
Solid (black) line: full asymmetry. Dashed (red) line: nonresonant asymmetry.
Flavor oscillations have been taken into account. 
}
\end{figure}

Figs. \ref{Fig:4.1} and \ref{Fig:4.2} display $A_{CP}(s)$ for $B_d\to \rho\mu^+\mu^-$ and 
$B_d\to \gamma\mu^+\mu^-$ decays, respectively. 

For $B_d\to \rho\mu^+\mu^-$ decays, the asymmetry reaches a 30-40\% level in the region of light vector resonances, and 
a level of 10\% between the light and $c\bar c$ resonances. Notice that flavor oscillations enhance the asymmetry by 
a factor 2.  

For $B_d\to \gamma\mu^+\mu^-$ decays the asymmetry is smaller and may be measured only in the region of light vector resonances. 

Both for $B_d\to \rho\mu^+\mu^-$ and $B_d\to \gamma\mu^+\mu^-$ decays the asymmetry is sensitive to the signs of the Wilson
coefficients $C_7$ and $C_9$. The asymmetry is however not sensitive to the invertion of the sign of $C_{10}$. 



\newpage

\subsubsection{Time-dependent asymmetry}
\begin{figure}
\begin{center}
\begin{tabular}{ccc}
\includegraphics[width=5.8cm]{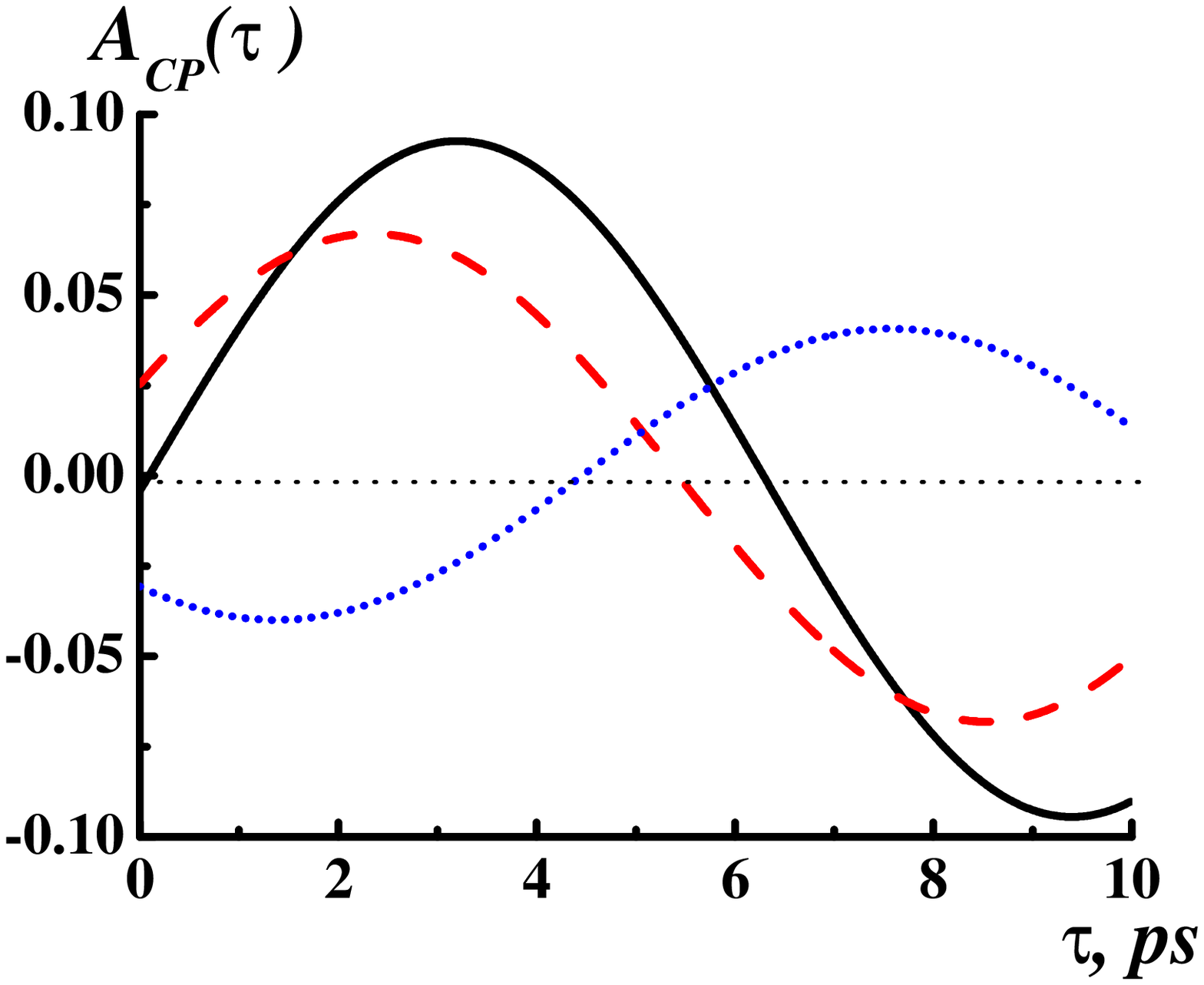} &
\includegraphics[width=5.8cm]{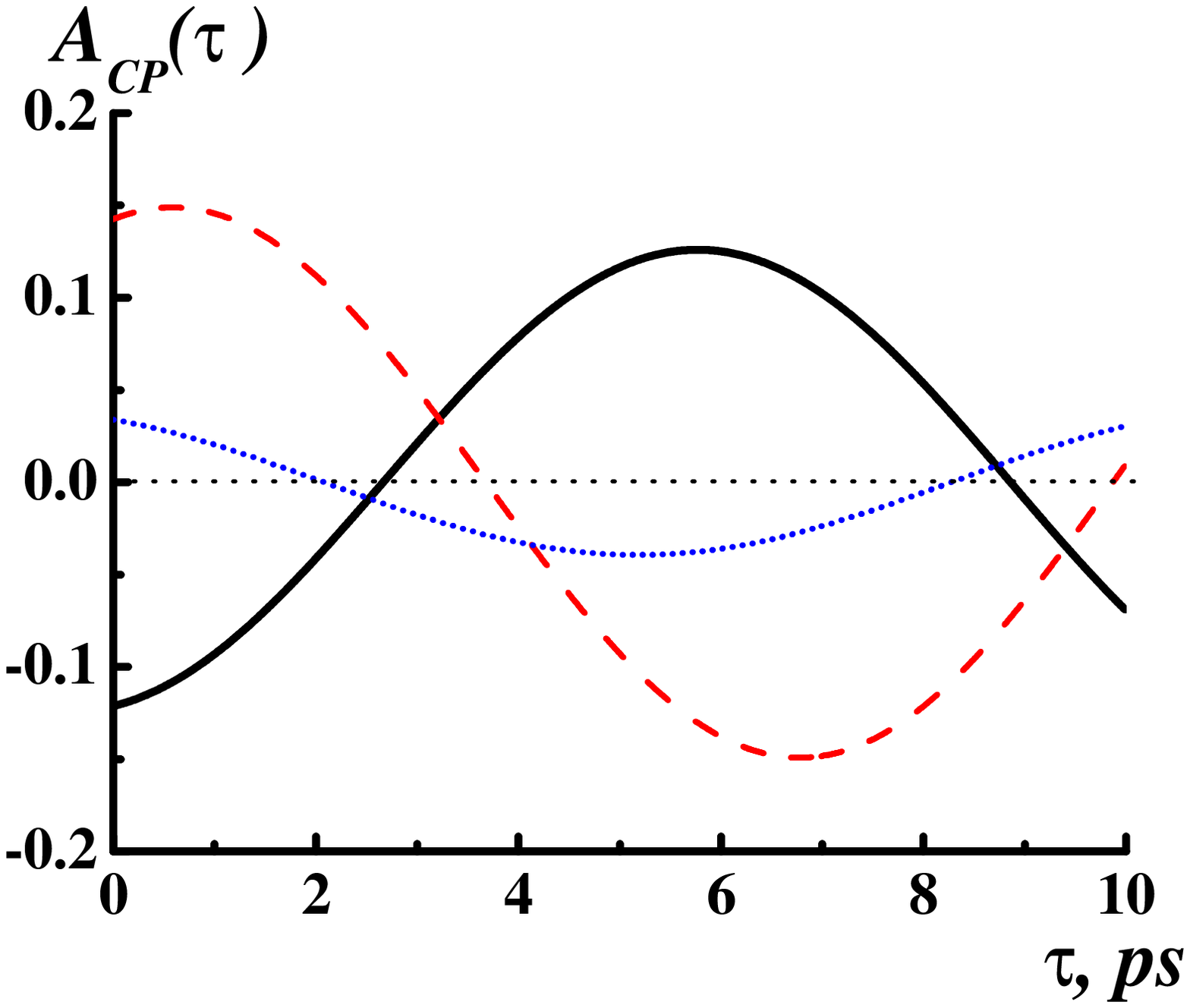} &
\includegraphics[width=5.8cm]{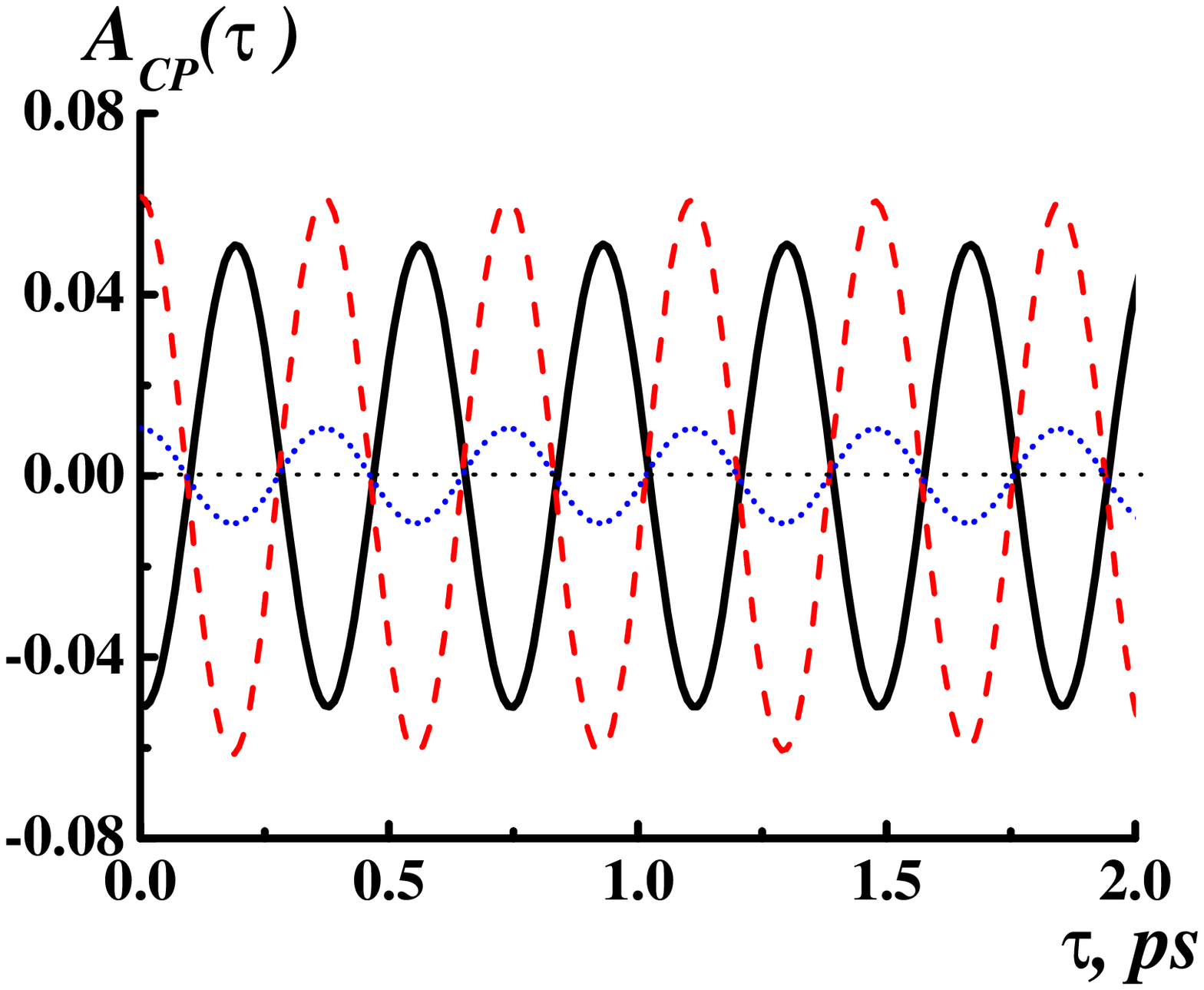} 
\\
(a) & (b) & (c)
\end{tabular}
\end{center}
\caption{\label{Fig:5}
Time-dependent asymmetry $A_{CP}(\tau)$: 
(a) $B_d\to \rho\mu^+\mu^-$; 
(b) $B_d\to \gamma\mu^+\mu^-$;  
(c) $B_s\to \gamma\mu^+\mu^-$. 
Solid line (black): SM. 
Dashed line (red): $C_{7\gamma}=-C^{\rm SM}_{7\gamma}$.  
Dotted line (blue): $C_{9V}=-C^{\rm SM}_{9V}$.}
\end{figure}
Fig.~\ref{Fig:5} plots the time-dependent asymmetry $A_{CP}(\tau)$ for 
$B_d\to \rho\mu^+\mu^-$ (a)  $B_d\to \gamma\mu^+\mu^-$ (b) and 
$B_s\to \gamma\mu^+\mu^-$ (c) decays. The region around the $J/\psi$ and $\psi'$ resonances $0.33\le \hat s \le 0.55$ 
was excluded from the integration 
while calculating the time-dependent asymmetries. This procedure corresponds to the analysis of the experimental data. 

The asymmetry in $B_d$ decays reaches a level of 10\% at the time-scale of a few $B$-meson lifetimes ($\tau_{B_d}$=1.53 ps) 
and may be studied experimentally. It also exhibits a sensitivity the the extentions of the SM. 


\section{Conclusions}
We presented the analysis of the forward-backward and the CP-violating asymmetries in rare semileptonic 
and radiative leptonic $B$-decays. Our results may be summarized as follows: 

\begin{enumerate}
\item
We obtained the analytic results for the time-dependent and time-independent $CP$-asymmetries in rare radiative 
leptonic $B$-decays $B_{d,s}\to \gamma \ell^+\ell^-$. 

\item 
We presented numerical results for the forward-backward asymmetry in $B_{s}\to \phi \mu^+\mu^-$ decays which may 
be measured in the near future at the LHCb. This asymmetry, as could be expected, has a very similar shape to the 
asymmetry in $B_{d}\to K^* \mu^+\mu^-$ decays and thus may be used for ``measuring'' the signs of the Wilson coefficients 
$C_{7\gamma}$, $C_{9V}$, and $C_{10A}$. 

\item 
We studied the forward-backward asymmetry in $B_{d,s}\to \gamma \ell^+\ell^-$ decays taking into account the vector resonance 
contributions, the Bremsstrahlung, and the weak annihilation effects. 
We noticed that the light neutral vector resonances strongly distort the shape of the asymmetry at 
small values of the dilepton invariant mass. In particular, in the SM these resonances lead to a sizeable shift 
of the zero point of the full asymmetry compared to the zero-point of the non-resonant asymmetry. 
The $A_{FB}$ in this reaction reaches 60\% and thus may be studied experimentally at the LHC and 
the future Super-B factory. 

\item
We analysed the CP-violating asymmetries (both time-dependent and time-independent) in $B_d\to\rho\mu^+\mu^-$, 
$B_s\to\phi\mu^+\mu^-$, and $B_{s,d}\to\gamma\mu^+\mu^-$ decays. 

The asymmetries in $B_s$ decays are found to be very small 
and therefore to be of no practical interest. 

The asymmetries in $B_d$ decays reach measurable values and thus might provide additional tests of the SM and its extentions. 
These potentially interesting cases are: 
(i) The time-independent CP-violating asymmetry $A_{CP}(\hat s)$ in 
$B_{d}\to\rho\mu^+\mu^-$ decays 
in the region below $c\bar c$ resonances (10-30 \% level) 
and $A_{CP}(\hat s)$ in $B_{d}\to \gamma\mu^+\mu^-$ in the region of 
light neutral vector resonances (5-10 \% level). 
(ii) The time-dependent CP-violating asymmetry $A_{CP}(\tau)$ in $B_{d}\to(\rho,\gamma)\mu^+\mu^-$ 
decays (10\% level). 
\end{enumerate}


\acknowledgments
We are grateful to S.~Baranov, A.~Berezhnoj, V.~Galkin, Yu.~Koreshkova, and W.~Lucha for discussions and to G.~Hiller 
for comments on the initial version of the paper. The work was supported in part by grant for leading scientific schools 
1456.2008.2, by FASI state contract 02.740.11.0244, and by FWF project P20573. 

\appendix

\section{\label{appendix_a}Helicity components of lepton currents}
Helicity components of the vector current 
$
j^{\mu}\,\left(\ell^-_{\lambda_2},\,\ell^+_{\lambda_1}\right)\,\equiv\,
\bar \ell (\vec k_2,\,\lambda_2 )\,\gamma^{\mu}\,\ell (\vec k_1,\,\lambda_1 )
$
take the form:
\begin{eqnarray}
j^{\mu} \,\left(\ell^-_R\,\ell^+_L\right)\,
           &=&\,2\,i\,
\varepsilon\,\left(0,\,\cos\theta\cos\varphi,+\,i\,\sin\varphi,\,\sin\varphi\cos\theta\,-i\,\cos\varphi,\,-\sin\theta\right);
\nonumber\\
j^{\mu}\left(\ell^-_L\ell^+_R\right)\,&=&\,2\,i\,\varepsilon\,
\left(0,\,-\,\cos{\theta}\cos{\varphi}\,+\,i\,\sin{\varphi},\,-\,\cos{\theta}\sin{\varphi}\,-\,i\,\cos{\varphi},\,\sin{\theta}\right);
\nonumber\\
j^{\mu} \,\left(\ell^-_R\,\ell^+_R\right)\,&=&\,-2\,i\,m\,\left(0,\,\sin{\theta}\cos{\varphi},\,\sin{\theta}\sin{\varphi},\,\cos{\theta}\right);
\nonumber\\
j^{\mu} \,\left(\ell^-_L\,\ell^+_L\right)\,&=&\,-2\,i\,m\,\left(0,\,\sin{\theta}\cos{\varphi},\,\sin{\theta}\sin{\varphi},\,\cos{\theta}\right).
\end{eqnarray}

For the axial current $a^{\mu}\,\left(\ell^-_{\lambda_2},\,\ell^+_{\lambda_1}\right)\,\equiv\,
\bar \ell (\vec k_2,\,\lambda_2 )\,\gamma^{\mu}\gamma^5\,\ell (\vec k_1,\,\lambda_1 )$
we find
\begin{eqnarray}
a^{\mu} \,\left(\ell^-_R\,\ell^+_L\right)&=&\,2\,i\, D\,\varepsilon\,\left(0,\,\cos{\theta}\cos{\varphi}\,+\,i\,\sin{\varphi},\,\cos{\theta}\sin{\varphi}\,-i\,\cos{\varphi},\,-\sin{\theta}\right);\nonumber\\
a^{\mu} \,\left(\ell^-_L\,\ell^+_R\right)&=&\,2\,i\, D\,
\varepsilon\,\left(0,\,\cos{\theta}\cos{\varphi}\,-\,i\,\sin{\varphi},\,\cos{\theta}\sin{\varphi}\,+\,i\,\cos{\varphi},\,-\sin{\theta}\right);\nonumber\\
a^{\mu} \,\left(\ell^-_R\,\ell^+_R\right)&=&\,-2\,i\,m\,\left(1,\,0,\,0,\,0\right);\nonumber\\
a^{\mu} \,\left(\ell^-_L\,\ell^+_L\right)&=&\,2\,i\,m\,\left(1,\,0,\,0,\,0\right).
\end{eqnarray}

For the tensor current
$
T^{\mu\nu}\,\left(\ell^-_{\lambda_2},\,\ell^+_{\lambda_1}\right)\,=\,
\bar \ell (\vec k_2,\,\lambda_2 )\,\sigma^{\mu\nu}\,\ell (\vec k_1,\,\lambda_1 )
$
we have:
\begin{eqnarray}
&&\frac{T^{\mu \,\nu} \,\left(\ell^-_R\,\ell^+_L\right)}{2\, m}\,
 =\,\left(
\begin {array}{cccc}
0                                              & -\cos{\theta}\cos{\varphi} - i\sin{\varphi} & -\cos{\theta}\sin{\varphi} + i\cos{\varphi} & \sin{\theta}\\
\cos{\theta}\cos{\varphi}\,+\,i\,\sin{\varphi} & 0 & 0 & 0\\
\cos{\theta}\sin{\varphi}\,-\,i\,\cos{\varphi} & 0 & 0 & 0\\
-\sin{\theta}                                  & 0 & 0 & 0\\
\end {array}
\right); \nonumber
\end{eqnarray}

\begin{eqnarray}
&&\frac{T^{\mu \,\nu} \,\left(\ell^-_L\,\ell^+_R\right)}{2\, m}\,=\,\left(
\begin {array}{cccc}
0                                               & \cos{\theta}\cos{\varphi} - i\sin{\varphi} & \cos{\theta}\sin{\varphi} + i\cos{\varphi} & -\sin{\theta}\\
-\cos{\theta}\cos{\varphi}\,+\,i\,\sin{\varphi} & 0 & 0 & 0\\
-\cos{\theta}\sin{\varphi}\,-\,i\,\cos{\varphi} & 0 & 0 & 0\\
\sin{\theta}                                    & 0 & 0 & 0\\
\end {array}
\right);\nonumber
\end{eqnarray}

\begin{eqnarray}
\frac{T^{\mu \,\nu} \,\left(\ell^-_R\,\ell^+_R\right)}{2\,\varepsilon}\,=\,
\left(
\begin {array}{cccc}
0                         & -\,\sin{\theta}\cos{\varphi}        & -\,\sin{\theta}\sin{\varphi}        & -\,\cos{\theta}                     \\
\sin{\theta}\cos{\varphi} &  0                                  &  i\, D\,\cos{\theta}                & -\,i\, D\,\sin{\theta}\sin{\varphi} \\
\sin{\theta}\sin{\varphi} & -\,i\, D\,\cos{\theta}              &  0                                  &  i\, D\,\sin{\theta}\cos{\varphi}   \\
\cos{\theta}              &    i\, D\,\sin{\theta}\sin{\varphi} & -\,i\, D\,\sin{\theta}\cos{\varphi} &  0                                  \\
\end {array}
\right);\nonumber
\end{eqnarray}

\begin{eqnarray}
\frac{T^{\mu \,\nu} \,\left(\ell^-_L\,\ell^+_L\right)}{2\,\varepsilon}\,=\,
\left(
\begin {array}{cccc}
0                             &  \sin{\theta}\cos{\varphi}            &  \sin{\theta}\sin{\varphi}          &  \cos{\theta}                       \\
-\,\sin{\theta}\cos{\varphi}  &  0                                    &  i\, D\,\cos{\theta}                & -\,i\, D\,\sin{\theta}\sin{\varphi} \\
-\,\sin {\theta}\sin{\varphi} & -\,i\, D\,\cos{\theta}                &  0                                  &  i\, D\,\sin{\theta}\cos{\varphi}   \\
-\,\cos{\theta}               &    i\, D\,\sin{\theta}\sin{\varphi}   & -\,i\, D\,\sin{\theta}\cos{\varphi} &  0                                  \\
\end {array}
\right).\nonumber
\end{eqnarray}


\section{Helicity amplitudes for rare semileptonic $\bar B(B)\to \bar V(V) \ell^+\ell$-decays \label{appb}}

\subsection{Kinematics}
We work in the rest frame of the $\ell^+\ell^-$-pair and choose z-axis in this reference frame along the 3-momentum of the $B$-meson; 
In this reference frame we have the following expressions for the 4-momenta of the initial meson ($p_1$),  the final meson ($p_2$), 
the negative-charged lepton ($k_1$), the positive-charged lepton ($k_2$), and the polarization vector of the final vector meson $\bar V (V)$: 
\begin{eqnarray}
\label{kin1}
&&k^{\mu}_1\, =\,\frac{M_1\,\sqrt{\hat s}}{2}\,\left (1,\, - D\,\vec n\, \right ),\qquad 
k^{\mu}_2\, =\,\frac{M_1\,\sqrt{\hat s}}{2}\,\left (1,\, D\,\vec n\, \right ),\nonumber\\
&&p^{\mu}_1\, =\,\left (E_1,\vec p_1\,\right )\, =\,\frac{M_1}{2\,\sqrt{\hat s}}\,
\left( 1 - \hat M_2^2 + \hat s,\, 0,\, 0,\,\hat\lambda^{1/2}\right ), \qquad
p^{\mu}_2\, =\,\left (E_2,\vec p_2\,\right )\, =\,\frac{M_1}{2\,\sqrt{\hat s}}\, 
\left( 1 - \hat M_2^2 - \hat s,\, 0,\, 0,\,\hat\lambda^{1/2}\right ), \nonumber\\
&&\epsilon^{*\,\mu}(\vec p_2,\,\lambda_V\, =\,\pm 1)\, =\, \frac{1}{\sqrt{2}}\, 
           \left (0,\,\mp 1,\, i,\, 0 \right ), \qquad
\epsilon^{*\,\mu}(\vec p_2,\,\lambda_V\, =\, 0)\, =\,
  \frac{1}{2\, \hat M_2\,\sqrt{\hat s}}\, 
  \left (\hat\lambda^{1/2},\, 0,\, 0,\, 1 - \hat M_2^2 - \hat s \right).  \nonumber 
\end{eqnarray}
where $\vec n\,=\,\left(\sin\theta\cos\varphi,\,\sin\theta\sin\varphi,\,\cos\theta\right)$, 
$s = (k_1 + k_2)^2$, $\hat s=s/M_1^2$, $\hat M_2=M_2/M_1$, $D = \sqrt{1\, -\, 4 \hat m^2 / \hat s}$, 
$\hat \lambda = \lambda \left (\hat s,\, 1,\, \hat M_2^2\right)$ 
with 
$\lambda (a, b, c)\, = a^2 + b^2 + c^2 - 2ab - 2bc - 2ac$. 

Three-particle phase space has the form 
\begin{eqnarray}
\label{ps3-cm-semilept}
d\Phi_3 &=&
\frac{(2\pi)^4\,\delta^4 (p_1-p_2-k_1-k_2)}
      {2 \varepsilon_1\, 2 \varepsilon_2\, 2 E_2}\, 
\frac{d\vec k_1}{(2\pi)^3}\, \frac{d\vec k_2}{(2\pi)^3}\,\frac{d\vec p_2}{(2\pi)^3}\, 
=\frac{M^2_1\,\hat \lambda^{1/2}}{2^8\,\pi^3}\, D\, d\hat s\, d\cos\theta .
\end{eqnarray}

\subsection{Helicity amplitudes}

The helicity amplitudes for the $\bar B_q$-decay ($q=s,d$)
\begin{eqnarray}
&&\bar A^{(q)}_{\lambda_V,\,\lambda_1,\,\lambda_2 }\, =\, 
\matrixel{\bar V_{\lambda_V}(p_2,\, M_2,\,\epsilon),\,\ell^+_{\lambda_1}(m,\, k_1),\,
          \ell^-_{\lambda_2}(m,\, k_2)}{H_{\rm eff}^{\rm SM}(b\to q\ell^+\ell^-)}
          {\bar B^0_q (M_1,\, p_1)}
\end{eqnarray}
read ($R_{qb}=\frac{G_F}{\sqrt{2}}\,\frac{\alpha_{\rm em}}{2\pi}\,V^*_{tq}\, V_{tb}$) : 
\begin{eqnarray}
\label{barB2Vll_matrix_el-1}
\bar A^{(q)}_{\pm 1,\, L,\, R} &=&
R_{qb}
M^2_1\, 
(1 \mp \cos\theta )\,\sqrt{\frac{\hat s}{2}}  
\left [   \pm\,\frac{\hat\lambda^{1/2}}{2}\, 
     \left (
       a\, +\, d\, D
     \right )\, -\, 
     \left (
       b\, +\, f\, D
     \right )
   \right ]. \nonumber\\
\label{barB2Vll_matrix_el-2}
\bar A^{(q)}_{\pm 1,\, R,\, L} &=&
R_{qb}M^2_1\, 
(1 \pm \cos\theta )\, \sqrt{\frac{\hat s}{2}} 
\left [
   \pm\,\frac{\hat\lambda^{1/2}}{2}\, 
     \left (
       a\, -\, d\, D
     \right )\, -\, 
     \left (
       b\, -\, f\, D
     \right )
   \right ]. \nonumber\\
\label{barB2Vll_matrix_el-34}
\bar A^{(q)}_{\pm 1,\, R,\, R} &=&\,\bar A^{(q)}_{\pm 1,\, L,\, L} =\, 
R_{qb}
M^2_1\, 
\sin\theta\,\frac{\hat m}{\sqrt{2}}\,  
 \left [ \hat\lambda^{1/2}\, a\, \mp\, 2 b \right ]. \nonumber\\
\label{barB2Vll_matrix_el-1-0}
\bar A^{(q)}_{0,\, L,\, R} &=& -\, 
R_{qb}
M^2_1\, 
\sin\theta\,\frac{1}{2 \hat M_2}\,
\left [
     \hat\lambda\, 
     \left (
       c\, +\, g\, D
     \right )\, -\,
     \left (1 - \hat M^2_2 - \hat s \right )\,  
     \left (
       b\, +\, f\, D
     \right )
   \right ]. \nonumber\\
\bar A^{(q)}_{0,\, R,\, L} &=&
R_{qb}
M^2_1\, 
\sin\theta\,\frac{1}{2 \hat M_2} \left[
     \hat\lambda\, 
     \left (
       c\, -\, g\, D
     \right )\, -\,
     \left (1 - \hat M^2_2 - \hat s \right )\,  
     \left (       b\, -\, f\, D \right )
   \right ]. 
\nonumber\\
\bar A^{(q)}_{0,\, R,\, R} \, &=&\,  -\, 
R_{qb}
 M^2_1\, 
\frac{\hat m}{\hat M_2 \sqrt{\hat s}}\,\hat\lambda^{1/2}
   \left [
     f - g (1 - \hat M^2_2) - 
     h\,\hat s + 
     \cos\theta\, 
     \left (
       \hat\lambda^{1/2}\, c\, -\, 
       \frac{1 - \hat M^2_2 - \hat s}{\hat\lambda^{1/2}}\, b
     \right ) 
   \right ]. \nonumber\\
\bar A^{(q)}_{0,\, L,\, L} \,& =&\,  
R_{qb} 
M^2_1
\frac{\hat m}{\hat M_2 \sqrt{\hat s}}\,\hat\lambda^{1/2}
 \left [
     f - g (1 - \hat M^2_2) - 
     h\,\hat s\, - 
     \cos\theta\, 
     \left (
       \hat\lambda^{1/2}\, c\, -\, 
       \frac{1 - \hat M^2_2 - \hat s}{\hat\lambda^{1/2}}\, b
     \right ) 
   \right ]. \nonumber
\end{eqnarray}
Here $\lambda_V$, $\lambda_1$ ($\lambda_2$) are the helicities of the vector meson and positive (negative) charged lepton, respectively. 

Similarly, the helicity amplitudes for the $B_q$-decay ($q=s,d$)
\begin{eqnarray}
&&A^{(q)}_{\lambda_V,\,\lambda_1,\,\lambda_2 }\, =\, 
\matrixel{V_{\lambda_V}(p_2,\, M_2,\,\epsilon),\,\ell^+_{\lambda_1}(m,\, k_1),\,
           \ell^-_{\lambda_2}(m,\, k_2)}{H_{\rm eff}^{\rm SM}(\bar b\to \bar q\ell^+\ell^-)}
          {B_q (M_1,\, p_1)} 
\end{eqnarray}
have the form ($R^*_{qb}=\frac{G_F}{\sqrt{2}}\,\frac{\alpha_{\rm em}}{2\pi}\,V_{tq}\, V^*_{tb}$)
\begin{eqnarray}
\label{B2Vll_matrix_el-1}
A^{(q)}_{\pm 1,\, L,\, R} &=&
R_{qb}^*
M^2_1\, 
(1 \mp \cos\theta )\, \sqrt{\frac{\hat s}{2}}  
\left [
   \mp\,\frac{\hat\lambda^{1/2}}{2}\, 
     \left (
       \tilde a\, +\, d\, D
     \right )\, -\, 
     \left (
       \tilde b +\, f\, D
     \right )
   \right ]. \nonumber\\
A^{(q)}_{\pm 1,\, R,\, L} &=&
R_{qb}^*
M^2_1\, 
(1 \pm \cos\theta )\,\sqrt{\frac{\hat s}{2}} 
\left [
   \mp\,\frac{\hat\lambda^{1/2}}{2}\, 
     \left (
       \tilde a -\, d\, D
     \right )\, -\, 
     \left (
       \tilde b\, -\, f\, D
     \right )
   \right ]. 
\nonumber\\
\label{B2Vll_matrix_el-34}
A^{(q)}_{\pm 1,\, R,\, R} &=& A^{(q)}_{\pm 1,\, L,\, L} = 
R_{qb}^*
M^2_1\, 
\sin\theta\, \frac{\hat m}{\sqrt{2}}\,  
 \left [ - \hat\lambda^{1/2}\, \tilde a\, \mp\, 2\tilde b \right ]. 
\nonumber\\
\label{B2Vll_matrix_el-1-0}
A^{(q)}_{0,\, L,\, R} \, &=&\,  -\, 
R_{qb}^*
M^2_1\, 
\sin\theta\,\frac{1}{2 \hat M_2}\,\left [
     \hat\lambda\, 
     \left (
       \tilde c\, +\, g\, D
     \right )\, -\,
     \left (1 - \hat M^2_2 - \hat s \right )\,  
     \left (
       \tilde b\, +\, f\, D
     \right )
   \right ]. 
\nonumber\\
\label{B2Vll_matrix_el-2-0}
A^{(q)}_{0,\, R,\, L} \, &=&\, 
R_{qb}^*
M^2_1\, 
\sin\theta\,\frac{1}{2 \hat M_2}\,
\left [
     \hat\lambda\, 
     \left (
       \tilde c\, -\, g\, D
     \right )\, -\,
     \left (1 - \hat M^2_2 - \hat s \right )\,  
     \left (
       \tilde b\, -\, f\, D
     \right )
   \right ]. 
\nonumber\\
\label{B2Vll_matrix_el-3-0}
A^{(q)}_{0,\, R,\, R} \, &=&\,  -\, 
R_{qb}^*
M^2_1\, 
\frac{\hat m}{\hat M_2 \sqrt{\hat s}}\,\hat\lambda^{1/2}
\nonumber\\
&&\times\left [
     f - g (1 - \hat M^2_2) - 
     h\,\hat s\, + 
     \cos\theta\, 
     \left (
       \hat\lambda^{1/2}\, \tilde c\, -\, 
       \frac{1 - \hat M^2_2 - \hat s}{\hat\lambda^{1/2}}\, \tilde b
     \right ) 
   \right ]. 
\nonumber\\
\label{B2Vll_matrix_el-4-0}
A^{(q)}_{0,\, L,\, L} \, &=&\,  
R_{qb}^*
M^2_1\, 
\frac{\hat m}{\hat M_2 \sqrt{\hat s}}\,\hat\lambda^{1/2}
\nonumber\\
&&\times\left [
     f - g (1 - \hat M^2_2) - 
     h\,\hat s\, - 
     \cos\theta\, 
     \left (
       \hat\lambda^{1/2}\, \tilde c\, -\, 
       \frac{1 - \hat M^2_2 - \hat s}{\hat\lambda^{1/2}}\, \tilde b
     \right ) 
   \right ]. \nonumber
\end{eqnarray}
In these formulas 
\begin{eqnarray}
&& a(\mu ,\, s)\, =\,
    4\, C_{7\gamma}(\mu)\,\frac{\left (\hat m_b + \hat m_q \right )}{\hat s}\, T_1(s)\, +\, 
    2\, C_{9V}^{{\rm eff}\, (q)}(\mu,\, s)\,\frac{V(s)}{1 + \hat M_2}, 
   \nonumber \\
&& \tilde a(\mu ,\, s)\, =\,
    4\, C_{7\gamma}(\mu)\,\frac{\left (\hat m_b + \hat m_q \right )}{\hat s}\, T_1(s)\, +\, 
    2\, C_{9V}^{{\rm eff}\, (\bar q)}(\mu,\, s)\,\frac{V(s)}{1 + \hat M_2}, 
   \nonumber \\
&& b(\mu ,\, s)\, =\,\left (1 + \hat M_2 \right )\,
   \left (
    2\, C_{7\gamma}(\mu)\,\frac{\left (\hat m_b - \hat m_q \right )}{\hat s}\, 
        \left (1 - \hat M_2\right )T_2(s)\, +\,
    C_{9V}^{{\rm eff}\, (q)}(\mu,\, s)\, A_1(s)  
   \right ), \nonumber \\
&& \tilde b(\mu ,\, s)\, =\,\left (1 + \hat M_2 \right )\,
   \left (
    2\, C_{7\gamma}(\mu)\,\frac{\left (\hat m_b - \hat m_q \right )}{\hat s}\, 
        \left (1 - \hat M_2\right )T_2(s)\, +\,
    C_{9V}^{{\rm eff}\, (\bar q)}(\mu,\, s)\, A_1(s)  
   \right ), \nonumber \\
&& c(\mu ,\, s)\, =\,\frac{1}{1 - \hat M_2^2}\, 
   \Biggl(
    2\, C_{7\gamma}(\mu)\,\frac{\left (\hat m_b - \hat m_q \right )}{\hat s}\, 
        \left (1 - \hat M_2^2\right )T_2(s)\, + \nonumber \\
&& \qquad\qquad
   +\, 2\, C_{7\gamma}(\mu)\,\left (\hat m_b - \hat m_q \right )\, T_3(s)\, +\,
   C_{9V}^{{\rm eff}\, (q)}(\mu,\, s)\,\left (1 - \hat M_2\right )\, A_2(s)  
   \Biggr ), \nonumber \\
&& \tilde c(\mu ,\, s)\, =\,\frac{1}{1 - \hat M_2^2}\, 
   \Biggl(
    2\, C_{7\gamma}(\mu)\,\frac{\left (\hat m_b - \hat m_q \right )}{\hat s}\, 
        \left (1 - \hat M_2^2\right )T_2(s)\, + \nonumber \\
&& \qquad\qquad
   +\, 2\, C_{7\gamma}(\mu)\,\left (\hat m_b - \hat m_q \right )\, T_3(s)\, +\,
   C_{9V}^{{\rm eff}\, (\bar q)}(\mu,\, s)\,\left (1 - \hat M_2\right )\, A_2(s)  
   \Biggr ), \nonumber \\
&& d(\mu ,\, s)\, =\, 2\, C_{10A}(\mu)\,\frac{V(s)}{1 + \hat M_2},\nonumber \\
&& f(\mu ,\, s)\, =\, C_{10A}(\mu)\,\left (1 + \hat M_2\right )\, A_1(s),\quad
   g(\mu ,\, s)\, =\, C_{10A}(\mu)\,\frac{A_2(s)}{1 + \hat M_2}, \nonumber \\
&& h(\mu ,\, s)\, =\,\frac{C_{10A}(\mu)}{\hat s}\,
   \Biggl (
     \left (1 + \hat M_2\right )\, A_1(s)\, -\,
     \left (1 - \hat M_2\right )\, A_2(s)\, -\,
     2\,\hat M_2\, A_0(s) 
   \Biggr ),  \nonumber  
\end{eqnarray}



\section{\label{appc} Helicity amplitudes for rare radiative leptonic $\bar B(B)\to\gamma\ell^+\ell^-$-decays }


\subsection{Kinematics}
We work in the rest frame of the $\ell^+\ell^-$-pair and choose z-axis in this reference frame along the 3-momentum of the $B$-meson.  
In this reference frame we have the following expressions for the 4-momenta of the initial meson ($p$),  the final photon ($k$), 
the negative-charged lepton ($k_1$), the positive-charged lepton ($k_2$), and the photon polarization vector $\epsilon$: 
\begin{eqnarray}
\label{k1k2}
&&
k^{\mu}_1\, =\,\frac{M_1\,\sqrt{\hat s}}{2}\,\left (1,\, - D\,\vec n\, \right ),\qquad 
k^{\mu}_2\, =\,\frac{M_1\,\sqrt{\hat s}}{2}\,\left (1,\, D\,\vec n\, \right ), \qquad 
p^{\mu} = \left (E, 0, 0, \omega \right )\, =\,\frac{M_1}{2 \sqrt{\hat s}}\,
\left (1+\hat s,\, 0,\, 0,\, 1-\hat s\right), 
\nonumber\\
&&
k^{\mu}\, =\,\left (\omega,0,0,\omega\,\right ), \qquad 
\epsilon^{*\alpha}\,\left(\vec k,\,\lambda_{\gamma} = \pm 1\right)= \frac{1}{\sqrt 2}\,\left(0,\,\mp 1,\,i,\,0\right).
\end{eqnarray}
where $\vec n\,=\,\left(\sin\theta\cos\varphi,\,\sin\theta\sin\varphi,\,\cos\theta\right)$, 
$s = (k_1 + k_2)^2$, $D = \sqrt{1\, -\, 4 \hat m^2 / \hat s}$, 
$\omega = M_1 (1-\hat s)/2\sqrt{\hat s}$, and now $\hat\lambda^{1/2} = 1 - \hat s$.

Three-particle phase space reads 
\begin{eqnarray}
\label{ps3-cm}
d\Phi_3 &=&
\frac{(2\pi)^4\,\delta^4 (p - k - k_1 - k_2)}
      {2 \varepsilon_1\, 2 \varepsilon_2\, 2 \omega}\, 
\frac{d\vec k_1}{(2\pi)^3}\, \frac{d\vec k_2}{(2\pi)^3}\,\frac{d\vec k}{(2\pi)^3}\, =
\frac{M^2_1\left (1 - \hat s\right)}{2^8\,\pi^3}\, D\, d\hat s\, d\cos\theta, 
\end{eqnarray}


\subsection{Helicity amplitudes}
The helicitly amplitudes for the $\bar B_q$-decay ($q=s,d$)
\begin{eqnarray}
\bar A^{(q)}_{\lambda_{\gamma},\,\lambda_1,\,\lambda_2 }\, =\, 
\matrixel{\gamma(k,\,\lambda_{\gamma}),\,\ell^+_{\lambda_1}(m,\, k_1),\,
           \ell^-_{\lambda_2}(m,\, k_2)}{H_{\rm eff}^{\rm SM}( b\to q\ell^+\ell^-)}
          {\bar B_q (M_1,\, p)}
\end{eqnarray}
read 
($R_{qb}\equiv \frac{G_F}{\sqrt{2}}\frac{\alpha_{\rm em}}{2\pi}\, V^*_{tq} V_{tb}$)
\begin{eqnarray}
\label{barB0_matrix_el-1}
&& \bar A^{(q)}_{\pm 1,\, L,\, R}\,=\,\pm\, 
|e|R_{qb}
M^2_1\, 
\left (1\,\mp\,\cos\theta\right )\,\frac{1 - \hat s}{2\sqrt{2}}
\nonumber 
\\
&&\times\left[\frac{2\, \hat m_b}{\sqrt{\hat s}}\, C_{7\gamma} 
\left (F_{TV}^{b \to q} \mp F_{TA}^{b \to q} \right ) + 
\sqrt{\hat s} \left (C^{{\rm eff} \, (q)}_{9V} + D \, C_{10A}\right )
\left (F_V \mp F_A\right )\, +\,\frac{4 \hat m^2}{\sqrt{\hat s}}\,
\frac{1 - \hat s}{(\hat t - \hat m^2)(\hat u - \hat m^2)}\, 
C_{10A}\,\frac{f_{B_q}}{M_1}
\right ]. \nonumber
\\
\label{barB0_matrix_el-2}
&& \bar A^{(q)}_{\pm 1,\, R,\, L}\,=\,\pm\, 
|e|R_{qb}
M^2_1\, 
\left (1\,\pm\,\cos\theta\right )
\frac{1 - \hat s}{2\sqrt{2}}\,
\nonumber\\
&&\times\left[\frac{2\, \hat m_b}{\sqrt{\hat s}}\, C_{7\gamma} 
\left (F_{TV}^{b \to q} \mp\, F_{TA}^{b \to q} \right ) + 
\sqrt{\hat s} \left (C^{{\rm eff}\, (q)}_{9V} - D \, C_{10A}\right ) 
\left (F_V \mp F_A\right )\, +\frac{4 \hat m^2}{\sqrt{\hat s}}\,
\frac{1 - \hat s}{(\hat t - \hat m^2)(\hat u - \hat m^2)}\, 
C_{10A}\,\frac{f_{B_q}}{M_1}
\right ].\nonumber
\nonumber \\
\label{barB0_matrix_el-3}
&&\bar A^{(q)}_{\pm 1,\, R,\, R}\,=|e|R_{qb}
M^2_1\, 
\sin\theta\,\frac{1 - \hat s}{\sqrt{2}}\,\frac{\hat m}{\sqrt{\hat s}}\,
\nonumber \\
&&\times\left[\frac{2\, \hat m_b}{\sqrt{\hat s}}\, C_{7\gamma}\,
\left (F_{TV}^{b \to q} \mp\,F_{TA}^{b \to q} \right )\, +\,
\sqrt{\hat s}\, C^{{\rm eff}\, (q)}_{9V}\,\left (F_V \mp\, F_A\right )\,\pm  
\frac{\sqrt{\hat s}}{(\hat t - \hat m^2)(\hat u - \hat m^2)}\,
\Bigl ( D\, (1 + \hat s)\,\mp\, (1 - \hat s) \Bigr )\,  
C_{10A}\,\frac{f_{B_q}}{M_1}
\right ].\nonumber
\\
\label{barB0_matrix_el-4}
&&\bar A^{(q)}_{\pm 1,\, L,\, L}\,=\,
|e|R_{qb}
M^2_1\, 
\sin\theta\,\frac{1 - \hat s}{\sqrt{2}}\,\frac{\hat m}{\sqrt{\hat s}}\,
\nonumber \\
&&\times\left[\frac{2\, \hat m_b}{\sqrt{\hat s}}\, C_{7\gamma}\,
\left (F_{TV}^{b \to q} \mp\, F_{TA}^{b \to q} \right )\, +\,
\sqrt{\hat s}\, C^{{\rm eff}\, (q)}_{9V}\,\left (F_V \mp\, F_A\right )\,\pm
\frac{\sqrt{\hat s}}{(\hat t - \hat m^2)(\hat u - \hat m^2)}\,
\Bigl ( D\, (1 + \hat s)\,\pm\, (1 - \hat s) \Bigr )\,  
C_{10A}\,\frac{f_{B_q}}{M_1}
\right ].\nonumber
\end{eqnarray}
Similarly, the helicity amplitudes for the $B_q$-decay 
\begin{eqnarray}
A^{(q)}_{\lambda_{\gamma},\,\lambda_1,\,\lambda_2 }\, =\, 
\matrixel{\gamma(k,\,\lambda_{\gamma}),\,\ell^+_{\lambda_1}(m,\, k_1),\,
           \ell^-_{\lambda_2}(m,\, k_2)}{H_{\rm eff}^{\rm SM}( \bar b\to \bar q\ell^+\ell^-)}
          {B_q (M_1,\, p)}
\end{eqnarray}
have the form ($R_{qb}^*= \frac{G_F}{\sqrt{2}}\frac{\alpha_{\rm em}}{2\pi}\, V_{tq} V^*_{tb}$)
\begin{eqnarray}
\label{B0_matrix_el-1}
&& A^{(q)}_{\pm 1,\, L,\, R}\,=\,
\pm\,
|e|R_{qb}^*
M^2_1\, 
\left (1\, \mp\,\cos\theta\right )\frac{1 - \hat s}{2\sqrt{2}}\, 
\nonumber\\
&&\times\left[\frac{2\, \hat m_b}{\sqrt{\hat s}}\, C_{7\gamma}
\left (F_{TV}^{\bar b \to \bar q} \pm F_{TA}^{\bar b \to \bar q} \right ) + 
\sqrt{\hat s} \left (C^{{\rm eff}\, (\bar q)}_{9V} + D\, C_{10A}\right )
\left (F_V \pm F_A\right )\, +\,\frac{4 \hat m^2}{\sqrt{\hat s}}\,
\frac{1 - \hat s}{(\hat t - \hat m^2)(\hat u - \hat m^2)}\, 
C_{10A}\,\frac{f_{B_q}}{M_1}
\right ],
\nonumber\\
\label{B0_matrix_el-2}
&& A^{(q)}_{\pm 1,\, R,\, L}\,=\,
\pm\,
|e|R_{qb}^*
M^2_1\, 
\left (1\, \pm\,\cos\theta\right )\frac{1 - \hat s}{2\sqrt{2}}\,
\nonumber \\
&&\times\left[\frac{2\, \hat m_b}{\sqrt{\hat s}}\, C_{7\gamma}
\left (F_{TV}^{\bar b \to \bar q} \pm F_{TA}^{\bar b \to \bar q} \right ) + 
\sqrt{\hat s} \left (C^{{\rm eff}\, (\bar q)}_{9V} - D \, C_{10A}\right )
\left (F_V \pm F_A\right )\, +\,\frac{4 \hat m^2}{\sqrt{\hat s}}\,
\frac{1 - \hat s}{(\hat t - \hat m^2)(\hat u - \hat m^2)}\, 
C_{10A}\,\frac{f_{B_q}}{M_1}
\right ]. 
\nonumber\\
\label{B0_matrix_el-3}
&&
A^{(q)}_{\pm 1,\, R,\, R}\,=\,
|e|R_{qb}^* 
M^2_1\, 
\sin\theta\,\frac{1 - \hat s}{\sqrt{2}}\,\frac{\hat m}{\sqrt{\hat s}}\,
\nonumber \\
&&\times\left[\frac{2\, \hat m_b}{\sqrt{\hat s}}\, C_{7\gamma}\,
\left (F_{TV}^{\bar b \to \bar q} \pm F_{TA}^{\bar b \to \bar q} \right )\, + \,
\sqrt{\hat s}\, C^{{\rm eff}\, (\bar q)}_{9V}\,\left (F_V \pm\, F_A\right )\pm\, 
\frac{\sqrt{\hat s}}{(\hat t - \hat m^2)(\hat u - \hat m^2)}\,
\Bigl ( D\, (1 + \hat s)\,\mp\, (1 - \hat s) \Bigr )\,  
C_{10A}\,\frac{f_{B_q}}{M_1}
\right ]. 
\nonumber\\
\label{B0_matrix_el-4}
&&A^{(q)}_{\pm 1,\, L,\, L}\,=\,
|e|R_{qb}^*
M^2_1\, 
\sin\theta\,\frac{1 - \hat s}{\sqrt{2}}\,\frac{\hat m}{\sqrt{\hat s}}
\nonumber \\
&&\times\left[\frac{2\, \hat m_b}{\sqrt{\hat s}}\, C_{7\gamma}\,
\left (F_{TV}^{\bar b \to \bar q} \pm F_{TA}^{\bar b \to \bar q} \right )\, + \,
\sqrt{\hat s}\, C^{{\rm eff}\, (\bar q)}_{9V}\,\left (F_V \pm\, F_A\right )\,\pm
\frac{\sqrt{\hat s}}{(\hat t - \hat m^2)(\hat u - \hat m^2)}\,
\Bigl ( D\, (1 + \hat s)\,\pm\, (1 - \hat s) \Bigr )\,  
C_{10A}\,\frac{f_{B_q}}{M_1}
\right ]. 
\nonumber
\end{eqnarray}
In these formulas
\begin{eqnarray}
   F_{TV}^{b \to q}(q^2) &=& 
     \left (1\, +\,\frac{m_q}{m_b}\right )\, 
     \left (F_{TV}(q^2, 0)\, +\, F_{TV}(0, q^2) \right )\, - \, 
     \frac{16}{3}\,
     \left (
       \frac{V_{ub} V_{uq}^*}{V_{tb} V_{tq}^*}\, +\,
       \frac{V_{cb} V_{cq}^*}{V_{tb} V_{tq}^*}
     \right )\, 
     \frac{a_1}{C_{7\gamma}}\,\frac{f_{B_q}}{m_b}, \nonumber \\
   F_{TA}^{b \to q}(q^2) &=& 
     \left (1\, -\,\frac{m_q}{m_b}\right )\, 
     \left (F_{TA}(q^2, 0)\, +\, F_{TA}(0, q^2) \right ). \nonumber
\end{eqnarray}
and 
\begin{eqnarray}
   F_{TV}^{\bar b \to \bar q}(q^2) &=& 
     \left (1\, +\,\frac{m_q}{m_b}\right )\, 
     \left (F_{TV}(q^2, 0)\, +\, F_{TV}(0, q^2) \right )\, + \, 
     \frac{16}{3}\,
     \left (
       \frac{V_{ub}^* V_{uq}}{V_{tb}^* V_{tq}}\, +\,
       \frac{V_{cb}^* V_{cq}}{V_{tb}^* V_{tq}} 
     \right )\, 
     \frac{a_1}{C_{7\gamma}}\,\frac{f_{B_q}}{m_b}, \nonumber \\
   F_{TA}^{\bar b \to \bar q}(q^2) &=&  
     \left (1\, -\,\frac{m_q}{m_b}\right )\, 
     \left (F_{TA}(q^2, 0)\, +\, F_{TA}(0, q^2) \right ). \nonumber
\end{eqnarray}

\end{document}